# Instability of stratified two-phase flows in rectangular ducts

Alexander Gelfgat and Neima Brauner
*School of Mechanical Engineering, Faculty of Engineering, Tel Aviv University, Tel Aviv 6997801, Israel*

Abstract

The linear stability of stratified two-phase flows in rectangular ducts is studied numerically. The linear stability analysis takes into account all possible infinitesimal three-dimensional disturbances and is carried out by solution of the associated eigenproblem. The neutral stability boundary and the corresponding critical wave number are obtained for liquid – liquid and air – water systems. Depending on the problem parameters, the instability sets in owing to short, intermediate, of long wave most unstable perturbations. Patterns of the most unstable disturbances are reported and discussed. It is shown that the instability arises due to shear, or interfacial mechanisms. Effects of the surface tension and of width/height aspect ratio are also studied. The results support the premise that the stability analysis of stratified two-phase flow in the simpler geometry of two-infinite plates can provide a reasonable estimation of the conditions for which this flow pattern can be considered to be linearly stable.

## 1. Introduction

The present computational study examines instabilities of pressure gradient driven stratified two-phase flow in rectangular ducts. This flow configuration is a generic model for basic studies in multiphase flows, and is a basic flow pattern in horizontal and inclined gas-liquid and liquid-liquid flows in a gravitational field, where the lighter fluid flows above the heavier one. Such flows are an important part of various technical applications, e.g., heat exchangers and miniature heat sinks (e.g., Lu et al., 1994, Wilmarth & Ishii, 1994, Li et al., 1998, Coleman & Garimella, 1999, Xu et al, 1999, Zhao et al., 2006).

Experimental studies show that the stratified flow configuration can be stable only for certain operational conditions. The critical conditions are commonly depicted on flow pattern maps and constitute the transitional boundary from stratified-smooth to other two-phase flow configurations that are observed upon crossing this boundary (e.g., stratified-wavy, slugs or pseudo slugs, stratified with drops at the interface, see for example the literature review in Taitel and Barnea, 2015, Brauner, 2003).

Starting from the late 60th, the major part of the theoretical studies devoted to the two-phase flow instabilities dealt with gas-liquid systems. The destabilizing mechanisms were naturally associated with wind generated waves. The two pertinent well-known mechanisms are the Kelvin-Helmholtz (K-H) instability (Lord Kelvin, 1871, Lamb, 1932), and the so-called Jeffreys' sheltering mechanism (Jeffreys, 1925, 1926). However, application of the results of those classical theories to predict the onset of instability of stratified two-phase flow in channels requires introducing tunable parameters, which were found to be system dependent, and therefore cannot be considered as predictive tools. Possible routes for introducing the effect of viscosity into inviscid K-H analysis in case of two-phase channel flow were examined (e.g., Funda and Joseph (2001), Kim et al., (2011)), yet the effect of the shear stresses in the two fluids was not accounted for.

A more physically comprehensive approach that was followed in many later studies is to consider viscous flow and to use the framework of the Two-Fluid (TF) model for stability analysis. It is based on the 1D transient continuity and momentum equations of the two phases (e.g., Wu et al, 1987, Andritsos et al., 1989, Lin & Hanratty, 1991, Brauner & Moalem Maron, 1991, Barnea & Taitel, 1994). As the velocity profiles are not modelled, the application of the TF model requires closure relations (e.g., for the steady and wave induced wall and interfacial shear stresses, velocity profile shape factors (e.g., Baruner & Moalem Maron, 1991, Kushnir et al., 2014). When wave induced (wall and interfacial) shear stress components in phase with the wave slope were included in the closure relations, a generalized stability criterion was obtained. The latter combines the so-called viscous KH mechanism (due to the inertia terms of both of the phases) and the sheltering mechanisms (Brauner & Moalem Maron, 1994, Kushnir et al., 2014, 2017). While the TF model can in principle be applied to any conduit geometry for analyzing the stability of a variety of gas-liquid and liquid-liquid systems, the predictions are critically dependent on the reliability of the closure relations that are implemented (Brauner & Moalem Maron, 1993, Ullman et al, 2004). In the absence of exact theory, closure relations which are borrowed from single-phase flow and empirical correlations are used. Moreover, the stability analysis is restricted to long wave perturbations, which are an inherent assumption for the TF model. Further studies showed that this basic assumption severely limits the feasibility of the TF model approach to serve as a general predictive tool for the stratified flow stability limits (e.g., Barmak 2016a,b , Kushnir et al., 2017).

An alternative approach is to carry out a rigorous stability analysis in the simpler two-plate geometry, namely the two-layer plane Poiseuille flow (e.g., Charru & Fabre, 1994, Ó Náraigh et al., 2014, Kaffel & Riaz, 2015, Barmak et al., 2016a,b). This approach is helpful for better understanding of mechanisms involved in destabilization of stratified flows. The problem is reduced to Orr-Sommerfeld equation for the stream function disturbances, defined in each sublayer and coupled via boundary conditions that account also for possible interface deformation and capillary forces. Yih (1967), who conducted long wave analysis, was the first to show that the viscosity stratification alone can cause interfacial instability for an arbitrary low Reynolds number. He introduced the concept of interface mode of long waves, which can coincide with any streamline and is neutrally stable in single-fluid flow, but triggers instability when viscosity stratification exists. Hooper and Boyd (1983) showed that the interface may be unstable also to short-wavelength perturbations. It was also demonstrated that the effect of surface tension is always stabilizing, whereas a density difference may have either stabilizing or destabilizing effect. In following studies (e.g., Yiantsios & Higgins, 1988, Tilly & Davis, 1994) it was argued that instability triggered by the short-wavelength disturbance is essentially due to a shear perturbation mode of the Tollmien-Schlichting type, which is modified by the interfacial effects.

In our recent studies (Barmak et al., 2016a,b) comprehensive linear stability analyses of horizontal and inclined stratified flows in the two-plate geometry were conducted, while considering all possible wavenumbers. The formulation allows for the Squire transformation (Hesla et al, 1986, Barmak et al., 2017), which shows that two-dimensional disturbances are most critical and therefore are the only ones to be studied. It was shown that depending on the flow conditions, the critical perturbations (i.e., those responsible for the onset of instability) can originate mainly at the interface (so-called "interfacial modes of instability"), or in the bulk of one of the phases (i.e., "shear modes"). However, the analysis revealed that there is no definite correlation between the type of instability and the perturbation wavelength. In particular, long waves do not necessarily imply interfacial mode instability, and the classification to shear or interfacial mode can be made only based on examination of the pattern of the disturbance stream function. Obviously, also with shear-mode instability, interfacial waves are predicted to grow exponentially. In this sense, the critical conditions for the onset of turbulence in either of the phases coincide with the stratified smooth stability boundary, as observed in the experiments conducted by Charles and Lilleleht (1965) and Kao and Park (1972).

The stability results obtained for two-phase flows between infinite plates offered a better physical insight into the phenomenon of two-phase stratified flow instability, and the consequential flow pattern transition. They enabled some important observations concerning channel size (scale-up and scale down) issues and the stability limits compared to single phase flow of either of the phases. In horizontal flows, almost the entire stable region is located within the rectangle formed by lines of the critical superficial velocity of each of the phases derived using the critical *Re* for the single phase flow, although some stabilization compared to the single-phase flow can occur (Barmak et al., 2016a). Systems and conditions where a long-wave disturbance is the critical one for triggering instability were identified, for which the long-wave analytical solution (or the Two-Fluid model) can be conveniently applied for predicting the stable stratified flow boundaries. Under those conditions the role of the (viscous) K-H mechanism in triggering instability could be examined and exactly quantified compared to the 'sheltering' mechanism (Kushnir et al., 2014, 2017).

However, due to the simplified infinite plates geometry considered, only qualitative comparison of the results with data obtained in pipe flow is possible, and the implications to the stability of stratified two-phase flow in rectangular ducts or pipes are still questionable. One obvious difference is already noted when considering the critical Reynolds number in single phase flow. For single phase plane Poiseuille flow, the critical Reynolds number predicted by linear stability analysis is equal to 5772 (critical wavenumber, $k$=1.02 e.g., Orszag, 1971), whereas for flow in a rectangular duct the critical Reynolds number is beyond $10^4$ when the aspect ratio is below ~3 (Theofilis et al, 2004). While the linear stability of single-phase flow in rectangular ducts was extensively studied in the literature (e.g., Tatsumi & Yoshimura, 1990, Theofilis et al, 2004, Demyanko, 2011, Adachi, 2013), to the best of our knowledge, no attempt has been made to carry out a rigorous stability analysis of stratified two-phase flow in rectangular ducts.

The purpose of the current study is to formulate the model and computational framework that will enable examining the impact of the lateral walls on the linear stability of stratified two- phase flow in horizontal rectangular ducts. In particular, we are interested in examining to what extent results obtained for the flow between infinite plates can still provide useful information on the effect of the system parameters on the flow destabilization. The analysis considers stability of a two-dimensional flow with respect to all possible three-dimensional infinitesimal disturbances without introducing any additional simplifications. Following our previous studies (Barmak et al, 2016a,b), we consider liquid-liquid and gas-liquid (air-water) as two representative systems . The stability boundaries are presented on the flow pattern map and are accompanied by critical wavenumbers and spatial patterns of the most unstable perturbations.

## 2. Formulation of the problem

### *2.1. General equations*

The configuration of stratified two-phase flow in a horizontal long rectangular duct of the height $H$ and width $L$ , where *A=L/H* is the aspect ratio, is sketched in Fig. 1. The flow is driven by an imposed pressure gradient acting in the *x*-direction. It is assumed that in a fully developed steady flow, the two immiscible fluids form a two-layer system, so that the volumetric flow rates in layers 1 and 2 (Fig. 1) attain prescribed values $Q_1$ and $Q_2$, respectively. Similarly to the stratified two-phase flow between two infinite plates (Ullmann et al, 2003), for specified two-phase system (i.e., fluids' properties and channel size), the layer heights $h_1$ and $h_2$ are defined by the volumetric flow rate ratio, $q_{12} = Q_1/Q_2$, and should be derived along with the velocity profile.

In the general case the dimensionless flow is described by velocity $\boldsymbol{u}_j = \left(u_j, v_j, w_j\right)$ and pressure $p_j$, $j$ =1 or 2, of the liquids 1 and 2, governed by the momentum and continuity equations defined into each liquid:

$$\frac{\partial \boldsymbol{u}_j}{\partial t} + \left(\boldsymbol{u}_j \cdot \nabla\right)\boldsymbol{u}_j = -\frac{\rho_1}{\rho_{12}\rho_j}\nabla p_j + \frac{1}{Re}\frac{\rho_1}{\rho_{12}\rho_j}\frac{\mu_{12}\mu_j}{\mu_1}\Delta \boldsymbol{u}_j + \frac{1}{Fr}\boldsymbol{e}_z \ , \tag{1}$$

$$\nabla \cdot \boldsymbol{u}_j = 0, \tag{2}$$

with no-slip boundary conditions at the channel walls

$$\boldsymbol{u}_j(x, y = 0, z) = \boldsymbol{u}_j(x, y = A, z) = \boldsymbol{u}_j(x, y, z = 0) = \boldsymbol{u}_j(x, y, z = 1) = 0, \tag{3}$$

and continuity of velocities and viscous stresses at the interface

$$\boldsymbol{n} \cdot (\boldsymbol{T}_1 \cdot \boldsymbol{n}) = \boldsymbol{n} \cdot (\boldsymbol{T}_2 \cdot \boldsymbol{n}), \quad \boldsymbol{\tau}^{(1)} \cdot (\boldsymbol{T}_1 \cdot \boldsymbol{n}) = \boldsymbol{\tau}^{(1)} \cdot (\boldsymbol{T}_2 \cdot \boldsymbol{n}), \quad \boldsymbol{\tau}^{(2)} \cdot (\boldsymbol{T}_1 \cdot \boldsymbol{n}) = \boldsymbol{\tau}^{(2)} \cdot (\boldsymbol{T}_2 \cdot \boldsymbol{n}). \tag{4}$$

Here $t$ is time, $\rho_j$ and $\mu_j$ are density and viscosity of the fluids, $\boldsymbol{T}_j$ is the viscous stress tensor, $\boldsymbol{n}$ is the unit normal to the interface, and $\boldsymbol{\tau}^{(1)}$ and $\boldsymbol{\tau}^{(2)}$ are the two unit vectors tangent to the interface. Choosing $H$ to be the length scale, and denoting the velocity scale (to be defined later) $V_0$, the time sale is $H/V_0$. The dimensionless governing parameters are the densities and viscosities ratios $\rho_{12} = \rho_1/\rho_2$ and $\mu_{12} = \mu_1/\mu_2$, the Reynolds number $Re = V_0 H \rho_2/\mu_2$, the Weber number $We = V_0^2 H \rho_2/\sigma$, and the Froude number $Fr = V_0^2/gH$, where $g$ is the gravity acceleration.

The base flow (dimensional) velocity profiles $\widehat{\boldsymbol{U}}_j = \left(\widehat{U}_j(y,z), 0,0\right)$ should be derived analytically or numerically subject to the conditions:

$$\frac{1}{HL}\int_0^{\hat{h}_1}\int_0^L \widehat{U}_1 d\hat{y} d\hat{z} = U_{1s}, \quad \frac{1}{HL}\int_{\hat{h}_1}^{\hat{h}_2}\int_0^L \widehat{U}_2 d\hat{y} d\hat{z} = U_{2s}, \tag{5}$$

where $U_{js}$(=$Q_j/LH$ ) represents the phase superficial velocity (the dimensional variables are denoted by ^), and the corresponding superficial Reynolds number is $Re_{js} = U_{js} H \rho_j/\mu_j$

The velocity scale $V_0$ in Barmak et al (2016) and several earlier and later studies of flows in infinite fluid layers was chosen to be the base flow velocity at the interface. It is not convenient for the present case, because the base flow interface velocity is a function of $y$ (see below), and it becomes known only after the base flow is evaluated. For a solution-independent formulation we choose here $V_0 = U_m = (U_{1s} + U_{2s})$, where $U_m$ is the mixture velocity.

*2.2 Base flow*

The only non-zero component $\hat{U}_j(y,z)$ of the dimensional base flow is a solution of the momentum equations:

$$\frac{\partial^2 \hat{U}_j}{\partial \hat{y}^2} + \frac{\partial^2 \hat{U}_j}{\partial \hat{z}^2} = \frac{\hat{G}}{\mu_j} \tag{6}$$

where $\hat{G}$ is the pressure gradient driving the flow, which is equal in both layers. The dimensional problem is solved in a rectangle $0 \leq \hat{y} \leq H, 0 \leq \hat{z} \leq L$ subject to the no-slip boundary conditions at $\hat{y} = 0$ and $\hat{z} = 0, L$, and continuity of axial velocity and shear stresses across the phases' interface at $\hat{z} = \hat{h}_1$. Assuming that $\hat{h}_1$ and the pressure gradient are known, analytical solution for the velocity profile can be obtained (e.g., Tang & Himmelblau, 1963, Landman, 1991) :

$$\hat{U}_1 = \frac{4L^2}{\pi^3}\frac{\hat{G}}{\mu_1}\sum_{\substack{n=1\\odd}}^{\infty}\frac{\sin(y_n)}{n^3}\left\{1 - \frac{\gamma_n}{\mu_{12}}\frac{exp[(z_n-h_n)]-exp[-(z_n+h_n)]}{1+exp[-2h_n]} - \frac{exp[z_n-2h_n]+exp[-z_n]}{1+exp[-2h_n]}\right\} \tag{7.1}$$

$$\hat{U}_2 = \frac{4L^2}{\pi^3}\frac{\hat{G}}{\mu_2}\sum_{\substack{n=1\\odd}}^{\infty}\frac{\sin(y_n)}{n^3}\left\{1 + \frac{\gamma_n}{\mu_{12}}\frac{exp[-(z_n-h_n)]-exp[-(2H_n-z_n-h_n)]}{1+exp[-2(H_n-h_n)]} - \frac{exp[-(H_n-z_n)]+exp[-(H_n+z_n-2h_n)]}{1+exp[-2(H_n-h_n)]}\right\} \tag{7.2}$$

Where

$$H_n = \frac{n\pi H}{L};\ h_n = \frac{n\pi\hat{h}_1}{L};\ z_n = \frac{n\pi\hat{z}}{L};\ y_n = \frac{n\pi\hat{y}}{L} \tag{8.1}$$

and

$$\gamma_n = \frac{[1-\text{sech}(h_n)]-\mu_{12}[1-\text{sech}(H_n-h_n)]}{1/\mu_{12}tanh(h_n)+tanh(H_n-h_n)} \tag{8.2}$$

Integrating the velocity profiles Eqs.(7.1, 7.2) over the corresponding flow cross sections (Eq.( 5) yields:

$$U_{1s} = \frac{8L^3}{H\pi^5}\frac{\hat{G}}{\mu_1}\sum_{\substack{n=1\\odd}}^{\infty}\frac{1}{n^5}\left\{h_n - tanh(h_n) - \frac{\gamma_n}{\mu_{12}}[1-\text{sech}(h_n)\right\} \tag{9.1}$$

$$U_{2s} = \frac{8L^3}{H\pi^5}\frac{\hat{G}}{\mu_2}\sum_{\substack{n=1\\odd}}^{\infty}\frac{1}{n^2}\left\{H_n - h_n - \tanh(H_n-h_n) + \frac{\gamma_n}{\mu_{12}}[1-\text{sech}(H_n-h_n)]\right\} \tag{9.2}$$

The dimensionless height of the lower layer $h_1 = \hat{h}_1/H$ (referred to as the insitu holdup), is derived from the prescribed values of the viscosity ratio, $\mu_{12}$ and the ratio of the supercritical velocities:

$$\frac{U_{1s}}{U_{2s}} = q_{12} = \frac{1}{\mu_{12}}\frac{\sum_{\substack{n=1\\odd}}^{\infty}\frac{1}{n^5}\left\{h_n-tanh(h_n)-\frac{\gamma_n}{\mu_{12}}[1-\text{sech}(h_n)\right\}}{\sum_{\substack{n=1\\odd}}^{\infty}\frac{1}{n^5}\left\{H_n-h_n-\tanh(H_n-h_n)+\frac{\gamma_n}{\mu_{12}}[1-\text{sech}(H_n-h_n)]\right\}} \tag{9.3}$$

The holdup $h_1$ is obtained by a numerical solution of the Eq. (9.3) using the secant method. Then the pressure gradient $\hat{G}$ is determined so that the dimensional superficial velocities attain the prescribed values (Eqs. 9.1. or 9.2).

Alternatively, the base state can be evaluated numerically. In this case Eq. (6) is solved on a chosen grid for $\hat{G} = 1Pa$ and an initial guess of the holdup $h_1$. The initial guess is taken from the analytical solution for the infinite layers system, as in Barmak et al (2016). Then the holdup is varied until the prescribed superficial-velocity ratio is obtained. The secant method is applied at this stage. Upon finding the correct holdup, the pressure gradient is rescaled to bring the superficial velocities to their given values.

The possibility to use either analytical or numerical evaluation of the base state allows for an additional verification of the code. A series of numerical experiments showed that the use of the analytical solution causes certain numerical problems. Thus, it is necessary to sum more than 1000 series terms to reach convergence within 5-6 decimal places. The evaluation of the series terms requires calculations of the exponents having positive and negative arguments of very large magnitudes, which requires some additional numerical treatment. Besides that, it is very difficult to monitor the residual term over the whole flow region. Thus, it was found that the use of the numerical solution for the base state is not only much faster and simpler, but also more reliable.

An additional argument for the use of numerically obtained base flow is connected with the linearized stability problem discussed below. The linearization is evaluated in the mathematical neighborhood of the steady state solution of the non-linear problem, which is represented by a numerical model. The numerical solution for the base state is the correct steady state solution of the numerical model. Thus, its use makes the whole formulation more consistent.

### *2.3 Linearized stability problem*

For stability studies we assume that the base state is perturbed by infinitesimally small perturbation that are represented as $\tilde{\boldsymbol{v}}_j = \left[\tilde{u}_j(y,z), \tilde{v}_j(y,z), \tilde{w}_j(y,z)\right] exp(ikx + \lambda t)$, $\tilde{p}_j = \tilde{p}_j(y,z)\ exp(ikx + \lambda t)$. The perturbed interface is described as $\eta = h_1 + \tilde{\eta}(y) exp(ikx + \lambda t)$, where $\eta$ is the elevation of the interface above the level $z = 0$, and $\tilde{\eta}$ is the infinitely small amplitude of its perturbation. The linearized governing equations read

$$\lambda \tilde{\boldsymbol{v}}_j + (\boldsymbol{U}_j \cdot \nabla)\tilde{\boldsymbol{v}}_j + (\tilde{\boldsymbol{v}}_j \cdot \nabla)\boldsymbol{U}_j = -\frac{\rho_1}{\rho_{12}\rho_j}\nabla \tilde{p}_j + \frac{1}{Re}\frac{\rho_1}{\rho_{12}\rho_j}\frac{\mu_{12}\mu_j}{\mu_1}\Delta \tilde{\boldsymbol{v}}_j \quad , \tag{10}$$

$$\nabla \cdot \widetilde{\boldsymbol{v}}_j = 0. \tag{11}$$

The above equations must be solved together with no-slip boundary conditions at the duct walls, continuity of velocity and viscous stresses at the interface, and the kinematic condition for the interface perturbation. The whole set of the linearized boundary conditions (see also Barmak et al, 2019) are listed below. The no-slip conditions:

$$\widetilde{\boldsymbol{v}}_j(x, y = 0, z) = \widetilde{\boldsymbol{v}}_j(x, y = A, z) = \widetilde{\boldsymbol{v}}_j(x, y, z = 0) = \widetilde{\boldsymbol{v}}_j(x, y, z = 1) = 0\,. \tag{12}$$

The kinematic condition:

$$\lambda\tilde{\eta} = \widetilde{w}_1(y, z = h_1) + U_1(y, z = h_1)\frac{\partial\tilde{\eta}}{\partial x} \quad . \tag{13}$$

Continuity of velocities at the interface:

$$\widetilde{w}_1(y, z = h_1) = \widetilde{w}_2(y, z = h_1), \quad \tilde{v}_1(y, z = h_1) = \tilde{v}_2(y, z = h_1), \tag{14}$$

$$\tilde{u}_1(y, z = h_1) = \tilde{u}_2(y, z = h_1) + [U_2(y, z = h_1) - U_1(y, z = h_1)]\tilde{\eta} \tag{15}$$

Denoting a jump of a function over the interface by double square brackets, the continuity of the viscous stresses at the interface reads:

$$[\![\boldsymbol{n}\cdot(\boldsymbol{T}\cdot\boldsymbol{n})]\!]_{z=h_1} = \left[\tilde{p}_2 - \tilde{p}_1 + \frac{2}{Re}\left(\mu_{12}\frac{\partial\widetilde{w}_1}{\partial z} - \frac{\partial\widetilde{w}_2}{\partial z}\right) - \frac{1}{We}\left(\frac{\partial^2\tilde{\eta}}{\partial y^2} - k^2\tilde{\eta}\right) - \frac{1-\rho_{12}}{Fr}\tilde{\eta}\right]_{z=h_1} = 0 \tag{16}$$

$$[\![\boldsymbol{\tau}^{(x)}\cdot(\boldsymbol{T}\cdot\boldsymbol{n})]\!]_{z=h_1} =$$

$$\left[\left(\frac{\partial U_2}{\partial y} - \mu_{12}\frac{\partial U_1}{\partial y}\right)\frac{\partial\tilde{\eta}}{\partial y} - \left(\frac{\partial^2 U_2}{\partial z^2} - \mu_{12}\frac{\partial^2 U_1}{\partial z^2}\right)\tilde{\eta} + \left(\mu_{12}\frac{\partial\tilde{u}_1}{\partial z} - \frac{\partial\tilde{u}_2}{\partial z}\right) - ik(1-\mu_{12})\widetilde{w}_1\right]_{z=h_1} = 0 \tag{17}$$

$$[\![\boldsymbol{\tau}^{(y)}\cdot(\boldsymbol{T}\cdot\boldsymbol{n})]\!]_{z=h_1} = \left[ik\left(\frac{\partial U_2}{\partial y} - \mu_{12}\frac{\partial U_1}{\partial y}\right)\tilde{\eta} + \left(\mu_{12}\frac{\partial\tilde{v}_1}{\partial z} - \frac{\partial\tilde{v}_2}{\partial z}\right) - (1-\mu_{12})\frac{\partial\widetilde{w}_1}{\partial y}\right]_{z=h_1} = 0 \tag{18}$$

Equations (10)-(18) define the eigenvalue problem for the complex time increment $\lambda$ and the perturbation being the corresponding eigenvector. The flow is unstable if at least one eigenvalue with a positive real part exists. The eigenvalue with the largest real part and the corresponding eigenvector are referred to as the leading or the most unstable one. The instability threshold is defined as a set of governing parameters at which the leading eigenvalue crosses the imaginary axis, so that its real part turns from negative to positive.

## 3. Numerical method

The numerical approach is the same as in Gelfgat (2007) with an addition that allows for minimization of a critical parameter over the wave number, as is realized in Gelfgat (2020). The problem was solved on staggered grids using the finite volume method. The grid was stretched near the tube borders and at both sides of the interface using the same stretching function as in Gelfgat (2007):

$$y \to A\left[\frac{y}{A} - a_y \cdot sin\left(2\pi\frac{y}{A}\right)\right], 0 \le y \le A;\ z \to z - a_z \cdot sin(2\pi z), 0 \le z \le 1 \quad (19)$$

The base flow was calculated by applying the direct sparse solver to the discretized equation (6) and computation of the holdup by the secant method.

The discretized linear stability problem reduces to the generalized eigenvalue problem:

$$\lambda\mathbf{B}(\widetilde{\boldsymbol{v}},\tilde{p})^T = \mathbf{J}(\widetilde{\boldsymbol{v}},\tilde{p})^T \quad (20)$$

Here $\mathbf{J}$ is the Jacobian matrix that defines the r.h.s. of the linearized problem and $\mathbf{B}$ is a diagonal matrix whose diagonal elements corresponding to the time derivatives of $\widetilde{\boldsymbol{v}}$ are equal to one, while the elements corresponding to $\tilde{p}$ and the boundary conditions are zeros, so that $det\mathbf{B} = 0$. Thus, the generalized eigenproblem (19) cannot be transformed into a standard one.

As mentioned above, the examination of the base-flow stability for a given set of the governing parameters requires computation of the leading eigenvalue $\hat{\lambda}$ having the largest real part for all real wavenumbers $k$. The base flow is unstable when $\hat{\lambda} = \max_k\{Real[\lambda(k)]\} > 0$. The value of the wavenumber yielding the maximum of $Real[\lambda(k)] = 0$ is referred to as the critical wave number, denoted as $k_{cr}$. The imaginary part of the leading eigenvalue, $Im[\hat{\lambda}(k_{cr})]$, provides the oscillation frequency of a slightly supercritical flow state. It is called the critical frequency and is denoted as $\omega_{cr}$. The corresponding eigenvector of (20), i.e., the leading eigenvector, defines the most unstable perturbation of the base state. Its amplitude, to within multiplication by a constant, represents the amplitude of an oscillatory flow resulting from the onset of instability. Since all the disturbances at the critical point are proportional to $exp(ik_{cr}x + i\omega_{cr}t)$, they appear as a traveling wave propagating with the speed $c = -\omega_{cr}/k_{cr}$.

The eigenproblem (20) is solved by the Arnoldi iteration in the shift-and-invert mode:

$$(\mathbf{J} - \lambda_0\mathbf{B})^{-1}\mathbf{B}(\widetilde{\boldsymbol{v}},\tilde{p})^T = \vartheta(\widetilde{\boldsymbol{v}},\tilde{p})^T, \quad \vartheta = \frac{1}{\lambda-\lambda_0} \quad (20)$$

where $\lambda_0$ is a complex shift. The Arnoldi method realized in the ARPACK package of Lechouq et al. (1998) is used. Following Gelfgat (2007), the LU decomposition of the complex matrix $(\mathbf{J} - \lambda_0\mathbf{B})^{-1}$ is computed, so that the calculation of the next Krylov vector for the Arnoldi method is reduced to one backward and one forward substitution.

In the following calculations we fix the duct geometry, the physical properties of the fluids, and one of the superficial velocities $U_{1s}$ or $U_{2s}$. The other superficial velocity is varied until its critical value is found, for which the perturbations of all wave numbers are damped, except for $k_{cr}$, at which the perturbation is neutrally stable (i.e., $Re\left[\hat{\lambda}(k_{cr})\right]$=0). The goal of a series of such calculations is to build a stability diagram similar to those reported in Barmak et al (2016) for two-phase flows in the two-plate geometry (i.e., infinite aspect ratio), in order to examine the effect of the duct aspect ratio on the phases' flow rates range where stratified flow with a smooth interface can be stable.

For calculation of the leading eigenvalue $\hat{\lambda}$, it is necessary to choose the shift $\lambda_0$ close to $\hat{\lambda}$ and to calculate 10-20 eigenvalues with the largest absolute value. Since at a neutral stability point of a specified $k$ the real part of the leading eigenvalue is zero, we fix *Real*($\lambda_0$)=0 and vary *Im*($\lambda_0$) until the leading eigenvalue $\hat{\lambda}$ corresponding to the specified values of $U_{1s}$ , $U_{2s}$ and *k* is computed. Then, either $U_{1s}$ or $U_{2s}$ is varied until the neutral stability point, where the leading eigenvalue is $\hat{\lambda} = \left(0, Im(\hat{\lambda})\right)$ is identified. The value of *Im*($\lambda_0$) is then further varied to ensure that there is no other eigenvalue with a larger real part. Subsequently, we vary the wavenumber $k$ to find at which $k = k_{cr}$ the neutral stability takes place at the lowest value of the superficial velocity being varied. At this stage we apply the golden ratio algorithm (Kiefer, 1953). For a given set of the geometrical and material parameters, the result of the stability study is defined by the critical values of the superficial velocities, $k_{cr}$, $\omega_{cr}$ and the leading eigenvector. Using a standard single-processor PC, computation of a point on stability diagrams reported below may consume from 20 to 60 minutes for a grid having $400^2$ nodes. Clearly, this time increases with the grid refinement, and can be decreased if the computational process is parallelized.

## 4. Test calculations

At first we addressed the instability of a single-phase flow in a rectangular tube and reproduced the results reported in Table 2 of Theofilis et al (2004), indicating that the critical Reynolds number decreases with increasing the aspect ratio towards the value of *Re*=5772 ($k_{cr} = 1.02$) for the flow between two infinite plates ($A \to \infty$), and that for small aspect ratios ($A < 3$) the flow is linearly stable at very large Reynolds numbers exceeding $10^4$. We also carried out several numerical experiments with the two-phase flow in a square duct for the parameters studied in Barmak (2016) and varying the grid stretching. It was found that the fastest convergence is observed with $a_y = a_z = 0.06$.

Then we addressed the analytical solution of Kushnir et al (2014) for the long-wave ($k = 0$) instability of a two-phase flow between two infinite plates. We considered parameters corresponding to the last row of the Table 1 of Barmak et al (2016), fixed the wavenumber at $k = 10^{-3}$, and studied stability for $U_{2s} = 1\,m/s$, and varying $U_{1s}$. Three values of the aspect ratio, $A = 10, 20,$ and 30 were examined (see Table 1). It was found that four converged decimal digits of $U_{1s,cr}$ and $\omega_{cr}$ can be calculated already on the 400×200 grid for $A = 10$ and $20$. For $A = 30$ three correct decimal digits are obtained already at 400×200 grid, and four digits starting form 800×400 grid. The converged critical values for $A = 10, 20,$ and 30 are, $U_{1s,cr} = 0.9600, 0.7990,$ and $0.7648$, $\widehat{\omega}_{cr} = -0.2960, -0.2483,$ and $-0.2375\ rps$, respectively. It can be observed that with the increase of the aspect ratio, the critical values of $U_{1s,cr}$ and $\omega_{cr}$ approach the analytical result for the flow of two infinite liquid layers, which is $U_{1s,cr} = 0.7312\ m/sec\,, \widehat{\omega}_{cr} = -0.2246\,rps$.

The convergence for the square pipe configuration is not as fast as for strongly elongated ducts, in particular due to the boundary layers which develop near the lateral boundaries and affect both the base flow and the most unstable perturbations. Two examples of convergence studies for a liquid – liquid and air – water systems are presented in Tables 2 and 3. For the liquid – liquid system (Table 2), we arrive to two correct decimal digits using 200×200 grid. In the case of the air – water system this convergence is reached with the 600×600 grid. To converge to within the third decimal digit, one needs 600×600 or a finer grid in the liquid –liquid case, and at least 1000×1000 nodes for the air – water system. For the calculations below, we monitored the convergence and ensured that it is within at least two decimal digits, which is sufficient for the graphical representation and physical interpretation of the results.

## 5. Results

The stability of stratified two-phase flow in rectangular duct is governed by 7 dimensionless parameters. These include the channel aspect ratio, $A$, the fluids' viscosity and density ratios, $\mu_{12}, \rho_{12}$, the flow rate ratio, $q_{12}$ and the two-phase $Re = V_0 H \rho_2/\mu_2$, $Fr = V_0^2/gH$ and $We = V_0^2 H \rho_2/\sigma$ numbers ($V_0 = U_{1s} + U_{2s}$).

There are several mechanisms, which can be responsible for destabilization of the smooth interface solution. Basically, they can be classified to the shear flow instability and to the interfacial instability. Shear flow instability, originates near the duct walls and is encountered also in single-phase Poiseuille flow. It is related to amplification of short wavelength Tollmien-Schlichting waves in either of the phases, which result in transition to turbulent flow for Reynolds numbers higher than critical. Similarly to two-phase stratified flow in the two-plate geometry, the shear flow instability may first evolve either in the light layer or in the heavier layer, due to the shear exerted by the upper or lower wall, respectively (Barmak et al., 2016a). However, in the case of flow in a duct of a finite aspect-ratio, shear instability may first evolve in either of the layers due to the shear at the lateral walls.

The interfacial instability is associated with viscosity and/or density stratification (e.g., Yih, 1967; Kushnir et al., 2014). Such instability is viewed as a result of interaction of the flows in the two layers, which are connected through the velocity and viscous stresses boundary conditions at the interface. As a result, the flow can become unstable for lower flow rates (lower superficial Reynolds number) than in single phase flow. Viscosity stratification ($\mu_{12} \neq 1$) produces a discontinuity (jump) across the interface in the primary flow velocity gradient $U_j'$, which leads to energy transfer from the primary flow to the disturbed flow and causes the 'viscosity induced' instability. According to Hooper and Boyd (1983) and Boomkamp and Miesen (1996), this is the dominant mechanism for the so-called 'interfacial instability'. It should be emphasized that since the stability problem is linear, several independent destabilizing mechanisms may be observed, which will appear as distinct unstable eigenfunctions.

Due to the large number of dimensionless parameters, performing a complete parametric study of the linear stability problem is practically infeasible. Therefore, to examine the effect of the channel aspect ratio on the destabilization of the flow, we focus on two case studies considered previously in Barmak et al (2016a) for examining the linear stability of two-phase flow between infinite plates (i.e., $A \rightarrow \infty$). The first case study is a "liquid – liquid" system with $\rho_1 = 1000\, kg/m^3$, $\mu_1 = 0.25 \cdot 10^{-3}\, kg/m \cdot sec$, $\rho_2 = 800\, kg/m^3$, $\mu_2 = 0.5 \cdot 10^{-3}$ $kg/m \cdot sec$, $\sigma = 0.03\, N/m$, where the light phase is more viscous ($\mu_{12}$=0.5, $\rho_{12} = 1.25$) . The second case study is "air – water" system with $\rho_1 = 1000\, kg/m^3$, $\mu_1 = 10^{-3}\, kg/m \cdot sec$, $\rho_2 =$ $1\,kg/m^3$ , $\mu_2 = 1.8 \cdot 10^{-5}\,kg/m \cdot sec, \sigma = 0.072\,N/m$ . In this case the heavier phase is obviously more viscous, and the density and viscosity contrast between the phases is much larger than the in the liquid-liquid system ($\mu_{12}$=55, $\rho_{12} = 1000$) . The duct height is fixed at $H = 0.02m$.

*5.1. Dependence on the aspect ratio*

The effect of the aspect ratio is demonstrated for the liquid–liquid system fixing the superficial velocity of the upper phase at $U_{2s} = 0.05\,m/sec$, and varying the aspect ratio from 1 to 10. The corresponding stability diagram is shown in Fig. 2. We observe that the instability sets in due to three most unstable modes replacing each other with the variation of the aspect ratio. These modes are depicted in Fig. 3. The segments of the neutral curve corresponding to the different eigenmodes are shown by different colors in Fig. 2a, and correspond to the segmented curves in Fig. 6b. Within the linear stability model, the change from one eigenmode branch to another is abrupt and takes place at parameters where the corresponding eigenvalues simultaneously cross the imaginary axis. Clearly, in the fully non-linear case, the points of the modes exchange should be treated as co-dimension 2 Hopf-Hopf bifurcation points (see, e.g., Kuznetsov, 2000).

The simplest way to illustrate different eigenmodes, which, in fact, are the most unstable disturbances, is depicting their absolute values. These, to within multiplication by a constant, represent the oscillation amplitudes of slightly supercritical flows, and indicate in which parts of the flow region the instability sets in. The eigenmodes absolute values are presented in Fig. 3 along with the pattern of corresponding base flow. The orange lines in the figures indicate the location of the interface.

As shown in Fig. 2, the instability arises owing to the first mode for $1 \le A \le 2.4$, in which we observe boundary layers of $\tilde{u}$ and $\tilde{v}$ near the upper and lower walls, and a large perturbation of the vertical velocity $\tilde{w}$ in the duct center (see Fig. 3). When the aspect ratio exceeds the value of $\approx$ 2.4, this mode is replaced by the second one, which can be interpreted as a self-replication of mode 1 in a wider geometry. It remains the most unstable mode up to $A \approx 2.9$. Nevertheless, these two modes differ qualitatively, as observed from the different location of the perturbation amplitude maxima associated with these two modes, either at the symmetry midplane $y = A/2$, or at both sides of it. A closer look at the real and imaginary parts of the eigenvector reveals that the perturbations $\tilde{u}$ and $\tilde{w}$ of mode 1 are symmetric, while the perturbation of $\tilde{v}$ is antisymmetric. For mode 2 it is the opposite: the perturbations $\tilde{u}$ and $\tilde{w}$ are antisymmetric, while the perturbation of $\tilde{v}$ is symmetric. Thus, in the case of mode 1, the instability sets in preserving the mirror symmetry with respect to the midplane, while in the case of mode 2 this symmetry is expected to be broken. The local maxima of the modes amplitude near the lateral boundaries should also be noted, which indicate that their effect cannot be neglected, and that a finer resolution, e.g., stretching, near the boundaries is needed to resolve the instability correctly.

In a wider duct of $A > 2.9$, the instability sets in due to mode 3. To illustrate that the perturbation patterns remain similar with the growth of aspect ratio, this mode is plotted for $A = 5$ and $10$ (Fig. 3). Here we observe thin boundary layers of $\tilde{u}$ and $\tilde{v}$ near the central part of the upper and lower horizontal walls. The maximal amplitude of $\tilde{u}$ and $\tilde{v}$ are observed in the lower less viscous layer, close to the bottom wall, and again, a large perturbation of the vertical velocity $\tilde{w}$ in the duct center (Fig. 3). No significant perturbations are observed near the lateral boundaries. As it was observed for the mode 1, here the perturbations $\tilde{u}$ and $\tilde{w}$ are symmetric, and the perturbation of $\tilde{v}$ is antisymmetric, so that slightly supercritical states are expected to preserve the mirror symmetry.

To gain some more understanding in the changes of flow stability properties with increasing the aspect ratio, we calculated an additional stability diagram (Fig. 4), in which we varied both superficial velocities keeping their ratio constant at $q_{12} = 1$. It is worth noting that for a given duct aspect ratio, the holdup is defined solely by the flow rate ratio and is independent on the phase superficial velocities magnitude. Moreover, the dependence of the holdup on the aspect ratio, for a fixed $q_{12}$, is by the lateral walls effect only, and becomes rather mild already for $A$>1 (Fig. 9).. At lower aspect ratios $0.5 \leq A \leq 1$, where a noticeable flow stabilization is observed, this dependence becomes slightly stronger, but still remains moderate: $h_1$=0.421 and 0.451 for A=0.5 and 1, respectively. The three modes observed in Fig. 3 are observed also in Fig. 4, with an addition of another mode that becomes the most unstable in a relatively short interval $2.6 \leq A \leq 3.05$. To illustrate the similarity of the first three modes with the previous ones, we show the absolute values of $\tilde{u}$ as inserts in Fig. 4a. The absolute values of perturbations of the fourth mode also are included as inserts. This new mode 4 is quite similar to mode 2, in which the large scale structures are shifted towards the lateral boundaries, and a new smaller scale structure appears in the center. Similar to mode 2, the perturbations of $\tilde{u}$ and $\tilde{w}$ of mode 4 are symmetric, while the perturbation of $\tilde{v}$ is anti-symmetric.

To compare the relative amplitudes of the perturbed interface deformations, we scale the calculated values of $\eta$ by the $|\tilde{u}|_{max}$ of the corresponding eigenmode. The comparison is presented in Fig. 5. The absolute value of $\eta$ of mode 2 zeroes at the center due to antisymmetry of the perturbation of vertical velocity, $\tilde{w}$ . The perturbations of $\tilde{w}$ of modes 1, 3 and 4 are symmetric, whereby the maximal values of the interface deformation amplitude is located at the interface center. Note that the relative amplitude of the interface deformation grows with the aspect ratio (Fig. 5), so that the larger is the aspect ratio, the stronger interface deformations can be expected.

We tried also to examine whether one could expect limiting values for $A \to \infty$, and whether these values will be the same as those obtained for the flow between parallel plates. For this purpose, using the grid of 1000×200 nodes we calculated the critical parameters for $A = 50$ and 100 ($U_{2s} = 0.05\, m/sec$). The results for a fixed $U_{2s} = 0.05\, m/sec$ ($k_{cr}$ and $\omega_{cr}$ are dimensionless) are $U_{1s,cr} = 0.113\, m/sec$, $k_{cr} = 5.32$, and $\omega_{cr} = -4.28$ for $A = 50$, and $U_{1s,cr} = 0.0433\, m/sec$, $k_{cr} = 8.18$, and $\omega_{cr} = -9.28$ for $A = 100$. The corresponding result for flow between the infinite plates are $U_{1s,cr} = 0.0826\, m/sec$, $k_{cr} = 2.893$, and $\omega_{cr} = -6.635$. Based on those results we cannot arrive at any conclusion about the limiting critical values, and we cannot confirm that effects of the lateral boundaries disappear at $A \to \infty$.

To gain more insight in the above issue we carried out calculations with fixed superficial velocity ratio for $q_{12} = 0.2, 1,$ and 5, and duct aspect ratios 50 and 100. The results are summarized in Table 4. We observe that the critical values of the superficial velocities are close. At the same time the values of critical wavenumber and critical frequency exhibit a noticeable scatter, which does not allow us to make a definite conclusion about the asymptotic behavior at A → ∞.

### *5.2. Liquid – liquid system in a square duct*

The stability diagram for the liquid – liquid system in a square duct is shown in Fig. 6a. The results are presented in the coordinates of the fluids superficial velocity, which are commonly used in the two-phase flow literature for defining the stable region of the stratified-smooth flow pattern, and the other bounding flow patterns. The corresponding Reynolds numbers, $Re_{1s}$ and $Re_{2s}$, of the fluids are also provided. The stratified-smooth flow is stable below and on the left of the solid curves, which demark the neutral stability boundary, and is unstable above and to their right. Note that the neutral stability boundary may non-monotonically change with variation of the superficial velocities. An example is observed near $U_{2s} = 0.17\ m/sec$: with the increase of $U_{1s}$ the instability first sets in at $U_{1s,cr} \approx 0.0018\ m/sec$ owing to the mode 1, then the flow stabilizes at $U_{1s,cr} \approx 0.008\ m/sec$, and then again becomes unstable at $U_{1s,cr} \approx 0.026\ m/sec$ owing to the mode 2.

The results of Barmak (2016a) for a two-phase flow between two infinite plates are shown in Fig. 6a by a dash line. Comparing results of the two problems, we see that at small values of $U_{2s}$ the flow in the square duct remains stable over a larger range of $U_{1s}$ than that obtained in the model of two infinite plates. However, for a wide range of flow rates there is a general similarity between the critical superficial velocities, which are of the same order of magnitude in the two geometries. An exception is very low $U_{1s}$ values, which correspond to low holdups of the heavier and less viscous phase. Compared to the flow between two-plates, at small $U_{1s}$, the stable region of stratified flow in a square duct extends over a much wider range of $U_{2s}$. Hence, with thin layer of the less viscous phase, the flow stabilization by reducing the aspect ratio, obtained in single phase flow (Theofilis et al, 2004), is observed also here.

The corresponding critical values of the wavenumber and the oscillation frequency are reported in Fig. 6b. Also here, we observe three distinct eigenmodes of the linear stability problem that are responsible for the onset of instability at all the values of the superficial velocities considered. These modes are shown in Fig. 7 in the same way as in Fig. 3. Apparently, these perturbation modes are completely different from those depicted in Fig. 3, as well as those reported in Barmak et al (2016a), and show that perturbation grow sometimes near the duct boundaries and sometimes near the liquid – liquid interface.

In Fig. 7 we observe that mode 1 develops in the upper layer close to the interface. Only perturbations of the vertical velocity $\tilde{w}$ penetrate into the bulk of the upper layer, while perturbations of the two other velocities remain located close to the interface implying an interfacial mode of instability. It explains the effect of the thin heavy layer on the destabilization of the flow of the upper phase that occupies almost the whole duct cross section, which when flowing alone in the square duct (with $A$=1) would remain linearly stable at any $U_{2s}$. Mode 2, is associated with larger holdups of the lower layer, where the effect of the boundary layers on lateral walls becomes apparent. Nevertheless, it can also be characterized as an interfacial mode, as all the perturbations are located near the interface. The maximal amplitude of the vertical velocity disturbance is positioned exactly on the interface. In this case one can expect large interface deformations. For even larger holdups of the lower (and less viscous) layer, mode 3 develops in the lower layer, where the effect of the lateral walls becomes more pronounced. The pattern of this mode is quite similar to that of the mode 2. The qualitative difference is noticeably weaker perturbations in the upper layer. Examination of the real and imaginary parts of the eigenmodes shows that mode 1 is antisymmetric and breaks the mirror symmetry of the flow, while modes 2 and 3 are symmetric, so that the flow symmetry is preserved.

The interface perturbations of the three most unstable modes are presented in Fig. 8. The absolute value of $\eta$ of mode 1 zeroes at the center due to antisymmetry of the vertical velocity perturbation. The modes 2 and 3 are symmetric, and the maximal value of the interface deformation amplitude is located at the interface center. As expected, mode 2 produces a larger interface deformation due to the vertical velocity maximum located at the interface.

### *5.3.Air – water system*

The stability diagram for the air – water system is shown in Fig. 9a, and the associated critical wavenumber and wave speeds in Fig. 9b. Comparing with the results for a similar flow between two infinite plates (Barmak et al, 2016), we see that at small $U_{1s}$ (i.e., thin water layer) the presence of the lateral walls stabilizes the flow, whereby the smooth-stratified flow configuration remains stable over somewhat wider range of gas superficial velocities. The stabilization effect is less pronounced than in the liquid-liquid system studied above (Fig. 6), where small values of $U_{1s}$ correspond to thin layer of the less viscous fluid. On the other hand, at small $U_{2s}$, (<0.1 m/s, i.e., thin air layer) the corresponding $U_{1s,cr}$ in the square duct is not always larger than that obtained in the two-plate geometry. However, also for the studied air-water system, there is a general similarity between the stable regions obtained in the two geometries. In particular, it is interesting to note that in the region where instability is set at similar values of $U_{1s,cr}$ and $U_{2s,cr}$ the critical perturbation was found to be long waves (i.e., $k_{cr} = 0$ and $\omega_{cr} = 0$) in these two geometries. In fact, the range of $U_{2s,cr}$, where $k_{cr} = 0$ is similar in both geometries (see Barmak et al., 2016a). It was verified numerically that the limit $c = -(\omega_{cr}/k_{cr})_{k_{cr}\to 0}$ exists, and the corresponding wave speed is shown in Fig. 9b along with the wavenumbers and wave speeds of the other most unstable modes. In fact, since the long wave is a kinematic wave, its velocity can be derived analytically from the base flow solution (see Kushnir et al, (2017):

$$c = \left(\frac{\partial U_{1s}}{\partial h_1}\right)_{V_0=const} \tag{21}$$

which in the case of stratified flow in a rectangular duct yields:

$$c = \frac{1}{(1+q_{12})^2}\frac{\partial q_{12}}{\partial h_1} \tag{22}$$

where $q_{12}$ is given in Eq.(9.3). As shown in Fig. 9b, the wave speed is larger than the mixture velocity (i.e., $c > 1$) over the stability boundary, except at high air superficial velocities associated with mode 3 and 4, which results in thin water layers.

Additionally, we examined the stabilizing effect of the surface tension by increasing its value in 10 times, from 0.072 N/m to 0.72 N/m. As is seen from Fig. 9, we observe a noticeable stabilization, while the disturbances modes remain similar, exhibiting similar dependences of the critical wavenumber and wave speed, as well as similarity in the perturbation patterns shown in Figs. 10 – 12 and discussed below. Expectedly, the main effect of higher surface tension is a reduction of $k_{cr}$, since surface tension stabilizes the shorter waves. Worth noting is the stabilization effect of the surface tension also when $k_{cr} = 0$. It results from the curvature of the perturbed interface in the lateral direction, which does not exists in the two-plate model (see Fig. 13. below). Obviously, even when the value of $k_{cr}$ is similar in the square duct and the two-plate geometries, the corresponding patterns of the most unstable perturbations are qualitatively different owing to the lateral boundaries that leads to the spanwise non-uniformity of perturbations.

The perturbation patterns associated with the 4 different modes obtained in the square duct are plotted in Figs. 10 – 12. Since the holdup of configurations corresponding to modes 1 and 2 is almost equal to 1 (i.e., the water occupies almost the entire flow area), these modes are reported together with their upper part zoomed in. Mode 1 preserves the mirror symmetry, while mode 2 is anti-symmetric. An interesting observation here is that in spite of the very small thickness of the air layer, it plays a crucial role in the instability onset (Figs. 10 and 11), which follows from large perturbation amplitudes there. With the increase of the surface tension, this mode is replaced by a similar but symmetric one, having three spanwise distributed structures instead of two. The other three modes do not exhibit any qualitative change with the 10 times increase of the surface tension.

The boundary layers on the lateral walls appear to have insignificant effect on the perturbation pattern of modes 1, and 3. These two modes can be characterized as interfacial mode of instability. In this respect, the critical perturbations of those modes are similar to those obtained in the two-plate geometry, which in the similar range of superficial velocities were also characterized as interfacial mode of instability (Barmak et al., 2016a). The effect of the lateral walls is noticeable in mode 2, and becomes evident in mode 4 (Fig. 12), which preserves the mirror symmetry. The disturbance develops in the much thicker air layer. Similarly to the disturbance pattern in the two-plate geometry, as the water layer becomes thinner, the critical perturbation is shifted to the bulk of the gas layer and can be characterized as shear mode of instability

The amplitudes of the interface deformation (Fig. 13) of the mirror symmetric modes 1, 3 and 4 have maximum at the symmetry plane $y = 0.5$, which is the result of the modes' symmetry-preserving features. The interface deformation of the anti-symmetric mode 2 vanishes at the midplane, while exhibiting steep maxima near the lateral boundaries. Mode 1 that corresponds to very thin air layers exhibits the smallest relative deformation of the interface.

Referring to experimental flow pattern maps obtained for air-water flow in channels of a similar size (e.g., Taitel & Barnea, 2015) indicates that instability of thick water layers (i.e., mode 1 and 2) results in transition to elongated-bubble or slug flow. Instability of thin water layers (i.e., mode 4) results in transition from stratified-smooth to stratified-wavy flow. In the region of intermediate water holdups associated with mode 3, transition from stratified-smooth to pseudo-slugs was observed, implying that this transition is associated with long-wave instability.

**Conclusions**

Linear stability of two-phase stratified flow in rectangular horizontal ducts is studied numerically by a comprehensive approach involving evaluation of the steady base flow state and calculation of the leading eigenvalues of the linearized stability problem. The numerical approach is based on propositions of Gelfgat (2007) with the extension to uniform spatial direction along which the disturbances are assumed to be periodic Gelfgat (2020). Additionally, the infinitesimal perturbations of the capillary boundary separating two phases are taken into account.

Since the number of governing parameters is too large to be covered in a single study, the main attention was devoted to flows in a square duct. For this configuration we addressed a liquid – liquid system and a gas-liquid (air – water) system, for which the stability diagrams and patterns of the most unstable disturbances are reported. The modes with intermediate wavelengths (i.e., wavelength ~*H*) and shorter waves are the most unstable and responsible for the instability of the studied liquid-liquid and air-water system almost in the entire range of holdups. Exceptions are air-water flows, where with intermediate superficial velocities of the phases, the critical perturbations are long waves. Nevertheless, even for long wave perturbation, surface tension should not be ignored due to the spanwise variation of the interface curvature.

We also studied the effect of gradual elongation of the spanwise duct dimension (i.e., the aspect ratio) reaching the limit of two-phase stratified flow between infinite plates studied earlier by Barmak et al (2016a). We observed that the critical flow rates in the limiting case of the flow between two infinite plates can be reached asymptotically, although the effect of the duct lateral walls on the critical wave number does not vanish even for very large aspect ratios (e.g., A=100)

Patterns of the most unstable perturbations reveal that depending on the flow parameters the instability can set in owing to shear modes developing along either of horizontal walls, or along the lateral walls, or at the interface. Clearly, these three main instability mechanisms can interact, bringing to a variety of different instability mechanisms, and therefore to qualitatively different flow patterns in slightly and strongly supercritical regimes. The maximum interface deformation of the critical perturbation is not necessarily in the duct midplane, as the critical perturbation can correspond to an anti-symmetric mode, where the interface deformation vanishes at the midplane and exhibits maxima near the lateral boundaries. No correlation was found between the most unstable disturbance wavelength and the type of the instability.

A quite unexpected finding is that in contrast to single phase flow, a relatively small difference was found between the stable region of stratified two-phase flow in the square duct geometry and the flow between infinite plates. It appears that the very existence of the interface between the phases, and the associated discontinuity in the viscosity and/or density, overwhelm the stabilization effect of the lateral walls obtained in single phase flows. Also, for specified superficial velocities of the phases, the aspect ratio has a moderate effect on the location of the interface in the base flow solution. This suggests that that rough estimation of the stable stratified flow region and its variation with the system parameters can be obtained by using the much simpler model of the flow between two infinite plates.

Acknowledgments:

This research was supported by Israel Science Foundation (ISF) grant No 415/18 and was enabled in part by support provided by WestGrid (www.westgrid.ca) and Compute Canada (www.computecanada.ca).The authors are thankful to I. Barmak for providing additional calculations for two-phase flow between infinite pates.

## References

Adachi, T. 2013. Linear stability of flow in rectangular ducts in the vicinity of the critical aspect ratio, European Journal of Mechanics B/Fluids 41, 163–168.

Andritsos, N., Williams, L., Hanratty, T. J. 1989. Effect of liquid viscosity on the stratified-slug transition in horizontal pipe flow, Int. J. Multiphase Flow 15, 877-892.

Barmak, I., Gelfgat, A., Vitoshkin, H., Ullmann, A., and Brauner, N. 2016a. Stability of stratified two-phase flows in horizontal channels. Phys. Fluids 28, 044101.

Barmak, I., Gelfgat, A., Vitoshkin, H., Ullmann, A., Brauner, N. 2016b. Stability of stratified two-phase flows in inclined channels, Phys. Fluids 28, 084101-1-26.

Barmak, I., Gelfgat, A., Ullmann, A., Brauner, N. 2017. On the Squire's transformation for stratified two-phase flows in inclined channels, Int. J. Multiphase Flow 88, 142-151.

Barmak I., Gelfgat A., Ullmann A., and Brauner N. 2019. Non-modal stability analysis of stratified two-phase channel flows. Int. J. Multiphase flow 111, 122-139.

Barnea D., and Taitel, Y. 1993. Kelvin-Helmholtz Stability Criteria for Stratified Flow, Viscous Versus Non-Viscous (Inviscid) Approaches, Int. J. Multiphase flow 19, 639-649.

Brauner, N. 2003. Liquid-Liquid Two-Phase Flow Systems in Modeling and Experimentation in Two-Phase Flow Phenomena, CISM Udine, Ed. V Bertola, Springer-Verlag (2003).

Brauner, N., Moalem Maron, D. 1991. Analysis of stratified/nonstratified transitional boundaries in horizontal gas-liquid flows, Chem. Eng. Sci. 46, 1849-1859.

Brauner, N., Moalem Maron, D. 1993. The role of interfacial shear modelling in predicting the stability of stratified two-phase flow, Chem. Eng. Sci. 48, 2867-2879.

Brauner, N.., Moalem Maron, D. 1994. Dynamic Model for the Interfacial Shear as Closure Law in Two-Fluid Models, Nuclear Eng. and Design 149, 67-79.

Charles, M. E., Lilleleht, L. U. 1965. An experimental investigation of stability and interfacial waves in co-current flow of two liquids, J. Fluid Mech. 22, 217-224.

Charru, F., Fabre, J. 1994. Long waves at the interface between two viscous fluids, Phys. Fluids 6, 1223-1235.

Coleman, J.W., Garimella, S. 1999. Characterization of two phase flow patterns in small diameter round and rectangular tubes, Int. J. Heat and Mass Transfer 42 ,2869–2881 (1999).

Demyanko, K. V., Nechepurenko, Yu. M. 2011. Dependence of the linear stability of Poiseuille flows in a rectangular duct on the cross sectional aspect ratio, Doklady Physics 56, 531–533

Funada, T., Joseph, D.D. 2001 Viscous potential flow analysis of Kelvin–Helmholtz instability in a channel. J. of Fluid Mech., 445 263–283.

Gelfgat, A. Y. 2007. Three-dimensional instability of axisymmetric flows, solution of benchmark problems by a low-order finite volume method. Int. J. Numer. Meths. Fluids 54, 269-294.

Gelfgat, A. Y. 2020. A comparative study on instability of steady flows in helical pipes. Fluid Mech. Res., to appear.

Hesla, T. I., Pranckh, F. R., Preziosi, L. 1986. Squire's theorem for two stratified fluids, Phys. Fluids 29, 2808-2811.

Hooper, A. P., Boyd, W. S. G. 1983. Shear-flow instability at the interface between two viscous fluids, J. Fluid Mech. 128, 507-528.

Jeffreys, H. 1925. On the formation of water waves by wind, Proc. Roy. Soc. London A107, 189-206.

Jeffreys, H. 1926. On the formation of water waves by wind(second paper), Proc. Roy. Soc. London A110, 241-247 .

Kaffel, A., Riaz, A. 2015. Eigenspectra and mode coalescence of temporal instability in two-phase channel flow. Phys. Fluids 27, 042101.

Kiefer J. 1953. Sequential minimax search for a maximum. Proc. Amer. Math. Soc. 4, 502-506.

Kao, T.W., Park,C. 1972. Experimental investigations of the stability of channel flows. Part 2. Two-layered co-current flow in a rectangular channel, J. Fluid Mech. 52, 401-423.

Kim, H., Padrino, J.C., Joseph, D.D. 2011. Viscous effects on Kelvin–Helmholtz instability in a channel, J. Fluid Mech. 680, 398–416.

Kushnir, R., Segal, V., Ullmann, A., Brauner, N. 2014. Inclined two-layered stratified channel flows: Long wave stability analysis of multiple solution regions, Int. J. Multiphase Flow 62, 17-29.

Kushnir, R., Segal, V., Ullmann, A., Brauner, N. 2017. Closure relations effects on the prediction of the stratified two-phase flow stability via the two-fluid model, Int. J. Multiphase Flow 97, 78-93.

Kuznetsov, Y. A. 2000. Elements of Applied Bifurcation Theory. Appl Math. Sci. vol. 112, Springer, 2000.

Lamb, H., Hydrodynamics, Cambridge University Press, 6th ed., (1932).

Landman M. J. 1991. Non-unique holdup and pressure drop in two-phase stratified inclined pipe flow, Int. J. Multiphase Flow 17, 377-394.

Lechouq R B, Sorensen D C and Yang C 1998. ARPACK Users' Guide, Solution of Large Scale Eigenvalue Problems with Implicitly Restarted Arnoldi Methods (SIAM, Philadelphia).

Li, G., Guo, L., Chen, X. 1998. The behavior and characteristics of the interfacial waves in gas-liquid two-phase separated flow through downward inclined rectangular channel, J. Thermal Sci. 7, 29–36.

Lin, P. Y., Hanratty, T. J. 1986. Prediction of the initiation of slugs with linear stability theory, Int. J. Multiphase flow 12, 79-98.

Lord Kelvin, W.T. 1871. Hydrokinetic solutions and observations. Philosophical Magazine 42, 362–377.

Lu, Q., Suryanarayana, N. V. 1994. Interfacial waves with condensation of a vapor flowing inside a horizontal rectangular duct, Exp. Heat Transfer 7, 303-318.

Ó Náraigh, L., Valluri, P., Scott, D. M., Bethune, I., Spelt, P. D. M. 2014. Linear instability, nonlinear instability and ligament dynamics in three-dimensional laminar two-layer liquid-liquid flows, J. Fluid Mech. 750, 464-506.

Orszag, S. A. 1971. Accurate solution of the Orr-Sommerfeld stability equation, J. Fluid Mech. 50, 689-703.

Tilley, B. S., Davis, S. H., Bankoff, S. G. 1994 Linear stability of two-layer fluid flow in an inclined channel, Phys. Fluids 6, 3906–3922

Taitel, Y., Barnea , D. 2015 Encyclopedia of Two-Phase Heat Transfer and Flow I : Fundamentals and Methods, Volume 1: Modeling of Gas Liquid Flow in Pipes, Edited by: John R Thome, World Scientific (2015) https://doi.org/10.1142/9310

Tang, Y. P. and Himmelblau, D. M. 1963 Interphase mass transfer for laminar concurrent flow of carbon dioxide and water between parallel plates, AIChE J. 9, 630-635.

Tatsumi T., Yoshimura T. 1990 Stability of the laminar flow in a rectangular duct, J. Fluid Mech. 212, 437-449

Theofilis V., Duck P. W., and Owen J. 2004 Viscous linear stability analysis of rectangular duct and cavity flows, J. Fluid Mech. 505, 249-286.

Troniewski, L., Ulbrich, R. 1984 Two-phase gas-liquid flow in rectangular channels, Chem. Eng. Science 39, 751-765

Ullmann, A., Zamir, M., Gat, S., and Brauner, N. 2003 Multi-holdups in co-current stratified flow in inclined tubes, Int. J. Multiphase Flow 29, 1565–1581

Ullmann, A., Goldstien, A., Zamir, M., Brauner, N. 2004 Closure relations for the shear stresses in two-fluid models for stratified laminar flows, Int. J. Multiphase Flow 30, 877–900

Wilmarth, T., Ishii, M. 1994 Two-phase flow regimes in narrow rectangular vertical and horizontal channels, Int. J. Heat Mass Transfer 37, 1749-1758

Wu, H. L., Pots, B. F. M., Hollenberg. J. F., Meerhoff, R. 1987 in: Proceedings of the 3rd International conference on Multiphase Flow, The Hague, Netherlands, 13-21. May

Xu, J.L. , Cheng, P., Zhao, T.S. 1999 Gas-liquid two-phase flow regimes in rectangular channels with mini/micro gaps, Int. J. Multiphase Flow 25 ,411-432

Yiantsios, S. G., Higgins, B. G. 1988 Linear stability of plane Poiseuille flow of two superposed fluids, Phys. Fluids 31, 3225–3238

Yih, C. S. 1967 Instability due to viscosity stratification, J. Fluid Mech. 27, 337–352

Zhao, Y., Chen, G., Yuan,Q., 2006 Liquid-liquid two-phase flow patterns in a rectangular microchannel, AIChE Journal 52, 4052-4060

Table 1. Long-wave instability in shallow rectangular pipes with $A = H/L > 10$ and $k = 10^{-3}$. Parameters $U_{2s} = 1\, m/sec$, $\rho_{12} = 1.25$, $\mu_{12} = 0.5$ correspond to the case reported in the last row of Table 1 of Barmak et al (2016). The result for $A = \infty$: $U_{1s,cr} = 0.7312\ m/sec$, $\omega_{cr} = -0.2246\, rps$. Stretching $a_y = a_z = 0.06$.

| $N_y \times N_z$ | $A = 10$ | | $A = 20$ | | $A = 30$ | |
|---|---|---|---|---|---|---|
| | $U_{1s,cr}\ (m/sec)$ | $\omega_{cr}(rps)$ | $U_{1s,cr}\ (m/sec)$ | $\omega_{cr}(rps)$ | $U_{1s,cr}\ (m/sec)$ | $\omega_{cr}(rps)$ |
| 300×150 | 0.9600 | -0.2960 | 0.7971 | -0.2478 | 0.7642 | -0.2373 |
| 400×200 | 0.9600 | -0.2960 | 0.7990 | -0.2483 | 0.7645 | -0.2374 |
| 500×250 | 0.9600 | -0.2960 | 0.7990 | -0.2483 | 0.7646 | -0.2374 |
| 600×300 | 0.9600 | -0.2960 | 0.7990 | -0.2483 | 0.7647 | -0.2374 |
| 700×350 | 0.9600 | -0.2960 | 0.7990 | -0.2483 | 0.7647 | -0.2374 |
| 800×400 | 0.9600 | -0.2960 | 0.7990 | -0.2483 | 0.7647 | -0.2375 |
| 900×450 | 0.9600 | -0.2960 | 0.7990 | -0.2483 | 0.7648 | -0.2375 |
| 1000×500 | 0.9600 | -0.2960 | 0.7990 | -0.2483 | 0.7648 | -0.2375 |
| 1100×550 | 0.9600 | -0.2960 | 0.7990 | -0.2483 | 0.7648 | -0.2375 |
| 1200×600 | 0.9600 | -0.2960 | 0.7990 | -0.2483 | 0.7648 | -0.2375 |
| 1300×650 | 0.9600 | -0.2960 | 0.7990 | -0.2483 | 0.7648 | -0.2375 |
| 1400×700 | 0.9600 | -0.2960 | 0.7990 | -0.2483 | 0.7648 | -0.2375 |
| 1500×750 | 0.9600 | -0.2960 | 0.7990 | -0.2483 | 0.7648 | -0.2375 |

Table 2. Convergence of the critical parameters for liquid – liquid case with $U_{1s} = 0.1\, m/sec$ , $\rho_1 = 1000\, kg/m^3$, $\mu_1 = 0.25 \cdot 10^{-3}\, kg/m \cdot sec$ , $\rho_2 = 800\, kg/m^3$ , $\mu_2 = 0.5 \cdot 10^{-3}\, kg/m \cdot sec, \sigma = 0.03\, N/m \quad H = 0.02m$ . Stretching $a_y = a_z = 0.06$.

| $N_y$ | $N_z$ | $U_{2s,cr}$ | $k_{cr}$ | $\omega_{cr}$ |
|---|---|---|---|---|
| 100 | 100 | 0.05225 | 2.8259 | -2.5198 |
| 200 | 200 | 0.05176 | 2.8397 | -2.5402 |
| 300 | 300 | 0.05181 | 2.8391 | -2.5433 |
| 400 | 400 | 0.05191 | 2.8367 | -2.5426 |
| 500 | 500 | 0.05190 | 2.8355 | -2.5425 |
| 600 | 600 | 0.05198 | 2.8337 | -2.5416 |
| 700 | 700 | 0.05203 | 2.8336 | -2.5434 |
| 800 | 800 | 0.05208 | 2.8325 | -2.5429 |
| 900 | 900 | 0.05212 | 2.8305 | -2.5401 |
| 1000 | 1000 | 0.05215 | 2.8296 | -2.5394 |

Table 3. Convergence of the critical parameters for water – air case with $U_{2s} = 1\, m/sec$ , $\rho_1 = 1000\, kg/m^3$, $\mu_1 = 10^{-3}\, kg/m \cdot sec$ , $\rho_2 = 1\, kg/m^3$ , $\mu_2 = 1.8 \cdot 10^{-5}\, kg/m \cdot sec, \sigma = 0.072\, N/m \quad H = 0.02m$ . Stretching $a_y = a_z = 0.06$.

| $N_y$ | $N_z$ | $U_{1s,cr}$ | $k_{cr}$ | $\omega_{cr}$ |
|---|---|---|---|---|
| 100 | 100 | 0.04151 | 3.9529 | -3.2609 |
| 200 | 200 | 0.03879 | 3.9884 | -3.2289 |
| 300 | 300 | 0.03801 | 4.0105 | -3.2284 |
| 400 | 400 | 0.03769 | 4.0144 | -3.2239 |
| 500 | 500 | 0.03752 | 4.0222 | -3.2259 |
| 600 | 600 | 0.03741 | 4.0265 | -3.2268 |
| 700 | 700 | 0.03734 | 4.0244 | -3.2236 |
| 800 | 800 | 0.03729 | 4.0244 | -3.2224 |
| 900 | 900 | 0.03725 | 4.0222 | -3.2199 |
| 1000 | 1000 | 0.03723 | 4.0222 | -3.2192 |

Table 4. Critical parameters for very large aspect ratios and fixed superficial velocity ratio.

| $A$ | $q_{12}$ | $U_{1s,cr}$ | $U_{2s,cr}$ | $k_{cr}$ | $\omega_{cr}$ |
|---|---|---|---|---|---|
| 50 | 1 | 0.0584 | 0.0584 | 3.23 | -4.52 |
| 100 | 1 | 0.0586 | 0.0586 | 3.24 | -4.52 |
| ∞ | 1 | 0.0604 | 0.0604 | 3.21 | -5.12 |
| 50 | 0.2 | 0.0210 | 0.105 | 4.76 | -6.16 |
| 100 | 0.2 | 0.0218 | 0.109 | 4.71 | -6.08 |
| ∞ | 0.2 | 0.0201 | 0.1005 | 5.06 | -5.40 |
| 50 | 5 | 0.147 | 0.0293 | 4.023 | -23.9 |
| 100 | 5 | 0.133 | 0.0266 | 4.616 | -6.86 |
| ∞ | 5 | 0.124 | 0.0248 | 0.237 | -0.142 |

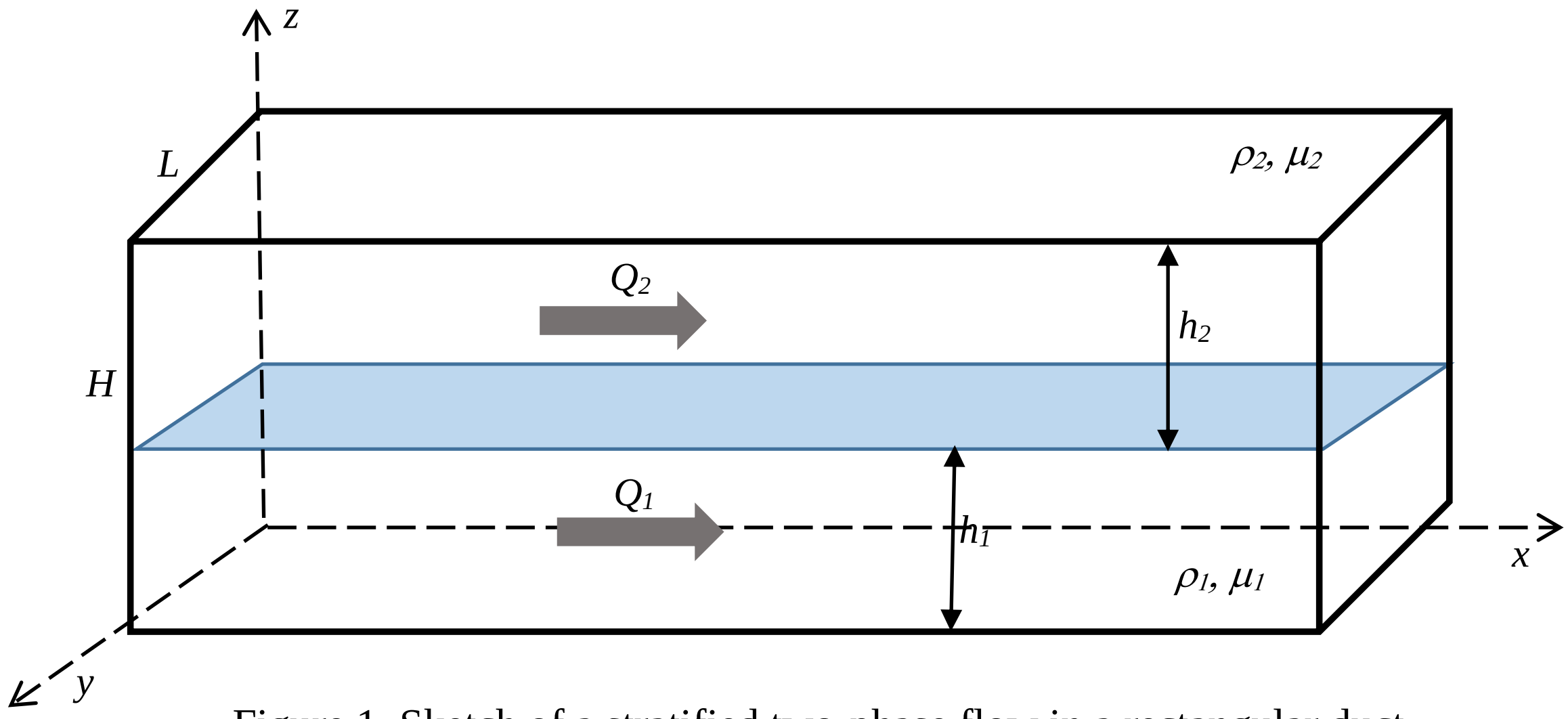

Figure 1. Sketch of a stratified two-phase flow in a rectangular duct.

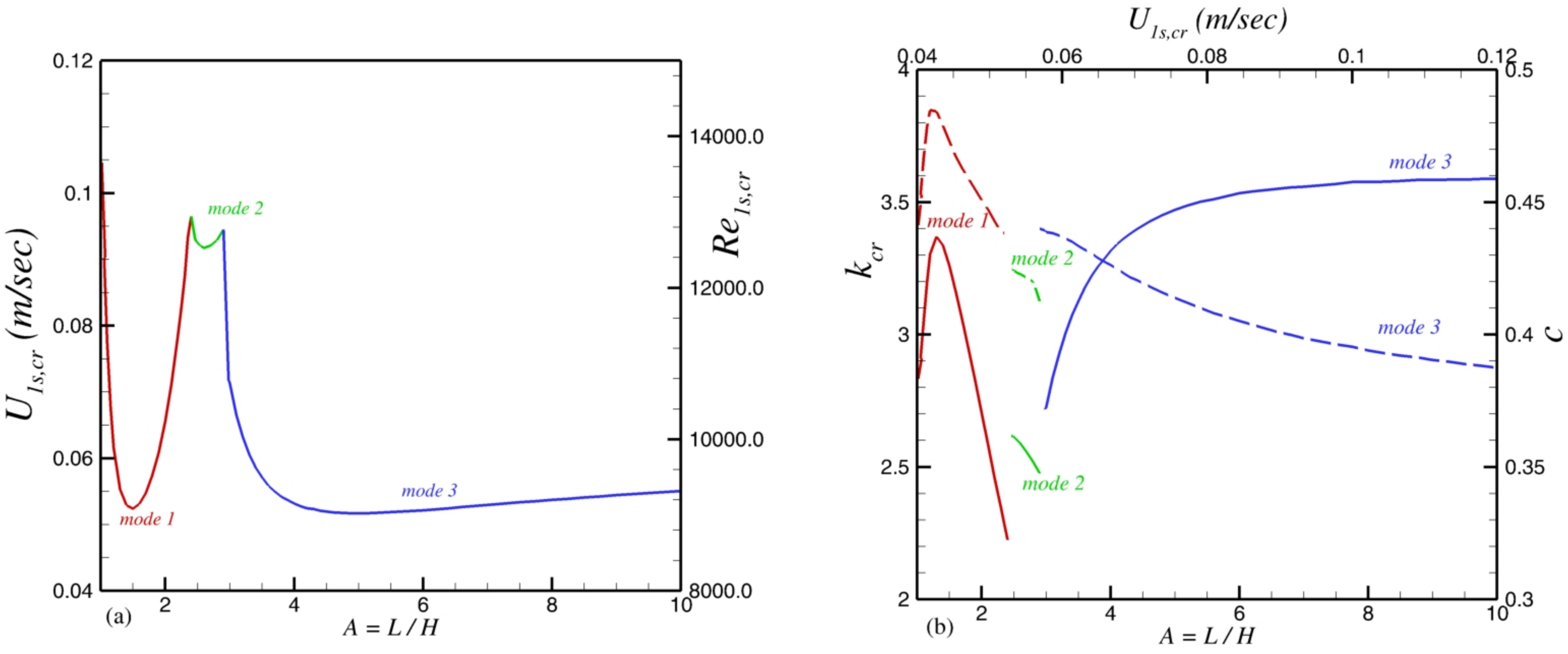

Fig. 2. Dependence of $U_{1s,cr}$ on the aspect ratio $A = L/H$ for a liquid-liquid system with fixed $U_{2s} = 0.05\, m/sec$. (a) Stability map and (b) variation of the dimensionless critical wavenumber (solid lines) and the wave speed (dash lines). The parameters of the liquid – liquid system are: $\rho_1 = 1000\, kg/m^3,\ \mu_1 = 0.25 \cdot 10^{-3}\, kg/m \cdot sec\, , \rho_2 = 800\, kg/m^3\, , \mu_2 = 0.5 \cdot 10^{-3}\, kg/m \cdot sec, \quad \sigma = 0.03\, N/m \quad H = 0.02m$ .

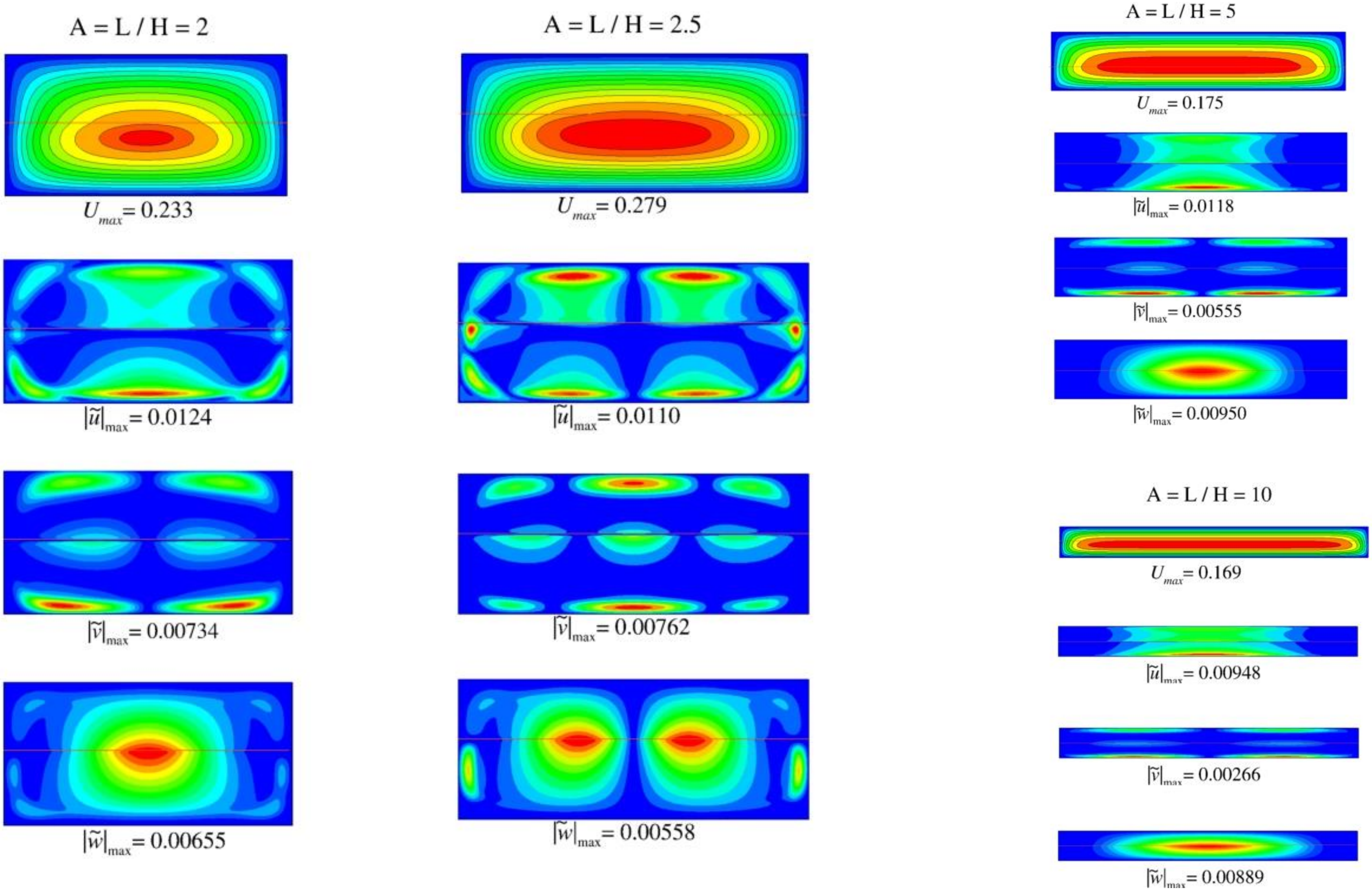

Fig. 3. The base flow (upper frames) and amplitudes of the velocity perturbations typical for the three instability modes depicted in Fig. 5.1.1. Mode 1: $A = 2, U_{1s} = 0.0655\, m/sec\ (q_{12} = 0.177),\ h_1 = 0.508$; Mode 2: $A = 2.5, U_{1s} = 0.0925\, m/sec\ (\ q_{12} = 0.185), h_1 = 0.564$; Mode 3: $A = 5, U_{1s} = 0.0517\, m/sec\ (q_{12} = 0.123),\ h_1 = 0.474$ and $A = 10, U_{1s} = 0.0550\, m/sec\ \ (q_{12} = 0.11),\ h_1 = 0.485\, m$.

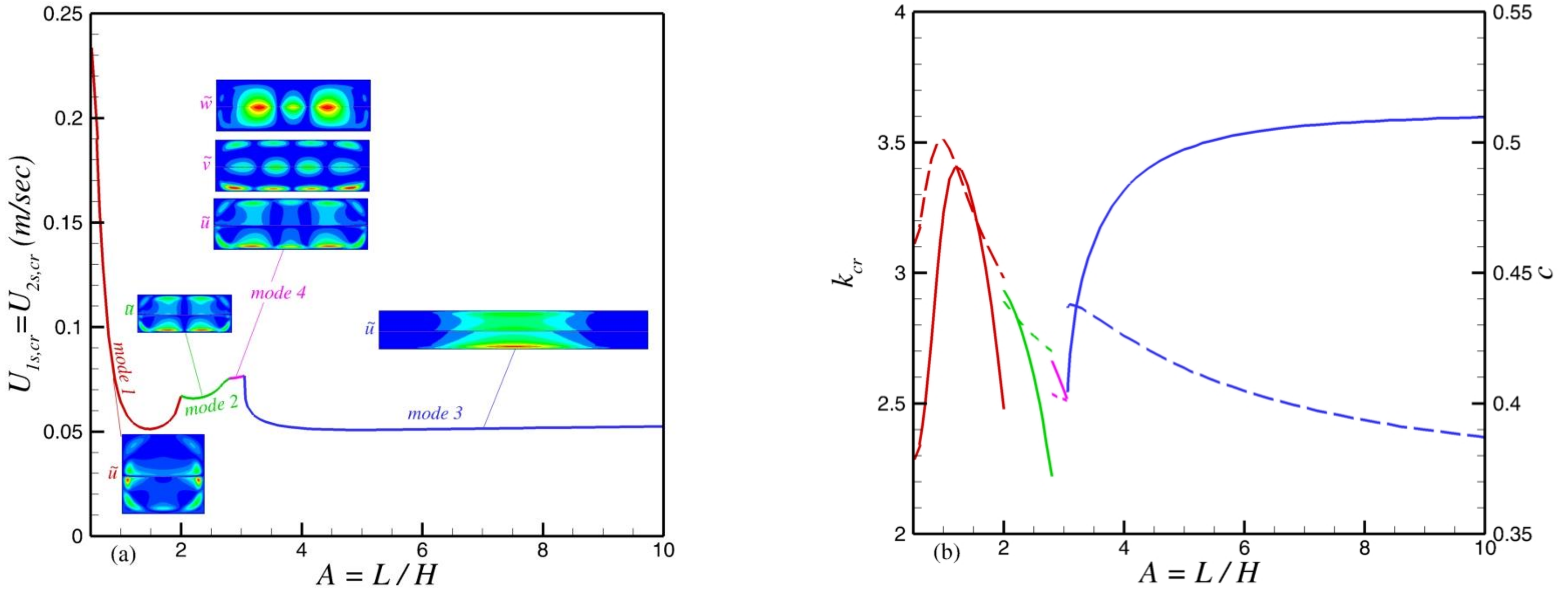

Fig. 4. Dependence of $U_{1s,cr} = U_{2s,cr}$ on the aspect ratio $A = L/H$ for a liquid-liquid system with fixed $q_{12} = 1$. (a) Stability map and (b) variation of the dimensionless critical wavenumber (solid lines) and the wave speed (dash lines). The parameters of the liquid – liquid system are: $\rho_1 = 1000\, kg/m^3$, $\mu_1 = 0.25 \cdot 10^{-3}\, kg/m \cdot sec$, $\rho_2 = 800\, kg/m^3$, $\mu_2 = 0.5 \cdot 10^{-3}\, kg/m \cdot sec$, $\sigma = 0.03\, N/m$ $H = 0.02m$ . The inserts of frame (a) show perturbation amplitudes of the streamwise velocity $\tilde{u}$ for the modes 1, 2, and 3, and the amplitudes of perturbations of all velocities for mode 4. Parameters for the inserts: Mode 1: $A = 1, U_{1s} = U_{2s} = 0.0651\, m/sec$, $h_1 = 0.451$; Mode 2: $A = 2.3, U_{1s} = 0.0661\, m/sec$, $h_1 = 0.465$; Mode 3: $A = 7, U_{1s} = 0.0514\, m/sec$, $h_1 = 0.469$; Mode 4: $A = 3, U_{1s} = 0.0762\, m/sec$, $h_1 = 0.466$

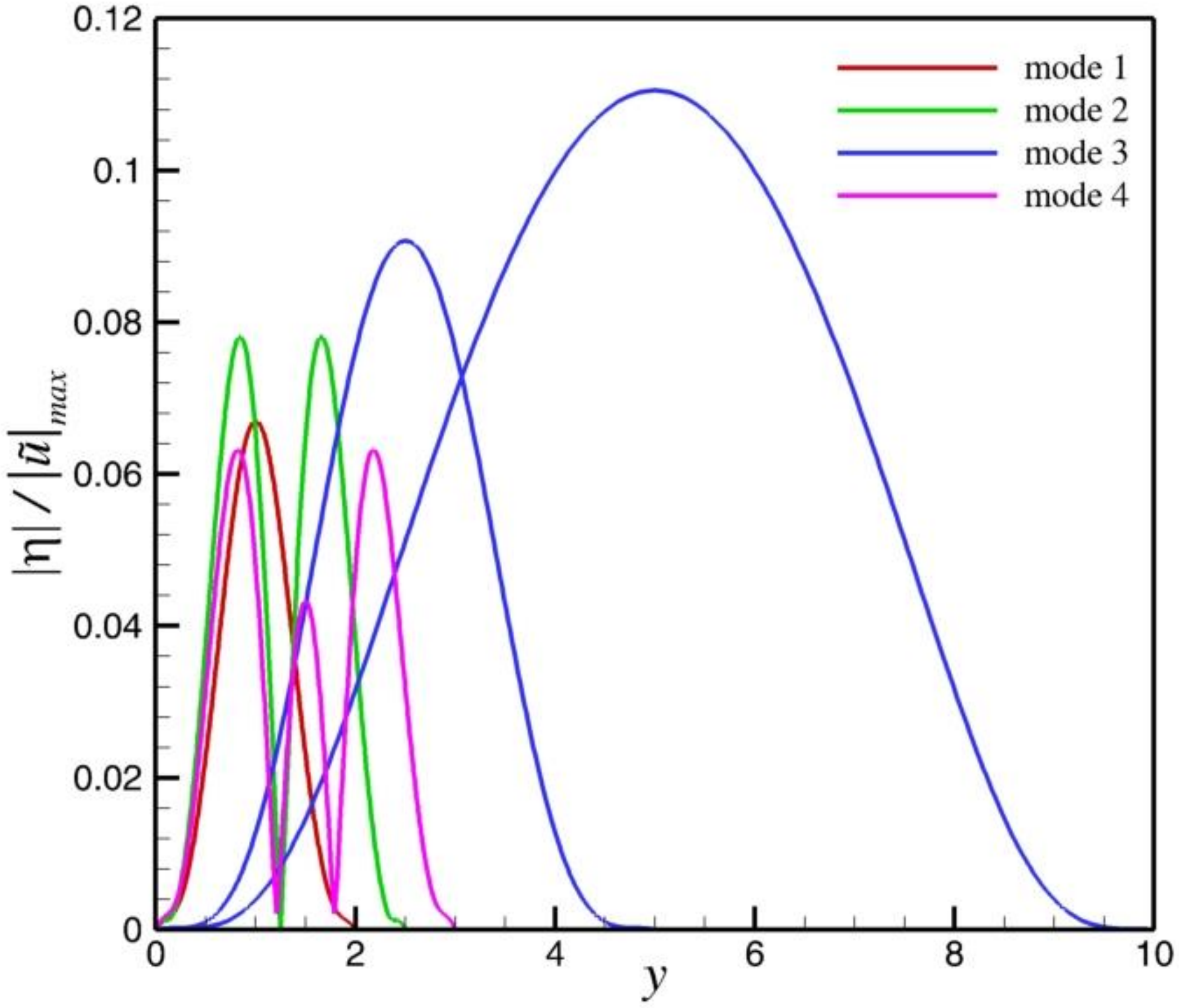

Fig. 5. Span-wise variation of the interface deformation amplitude for the three most unstable modes depicted in Fig. 5.1.2 (mode 1for $A$=2,  mode 2 for $A$=2.5, mode 3 for $A$=5 and 10, all for $U_{2s} = 0.05\, m/sec$; mode 4 for $A$=3 and $q_{12} = 1$).

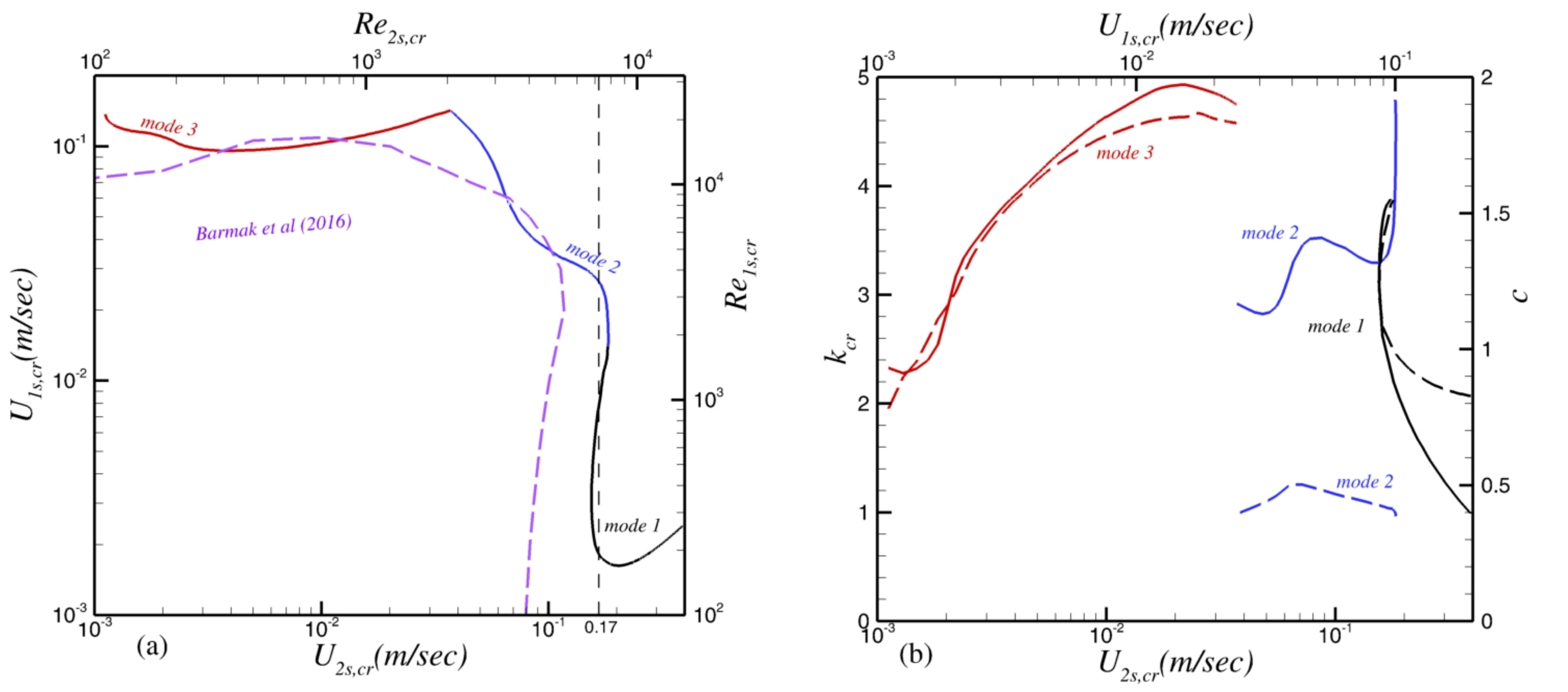

Fig. 6. (a) Stability map and (b) variation of the dimensionless critical wavenumber (solid lines) and the wave speed (dash lines) in a square channel ($H$=2cm , $A$=1) for a liquid – liquid system with $\rho_1 = 1000\,kg/m^3$, $\mu_1 = 0.25 \cdot 10^{-3}\,kg/m \cdot sec$, $\rho_2 = 800\,kg/m^3$, $\mu_2 = 0.5 \cdot 10^{-3}\,kg/m \cdot sec$, $\sigma = 0.03\,N/m$, $H = 0.02m$ .

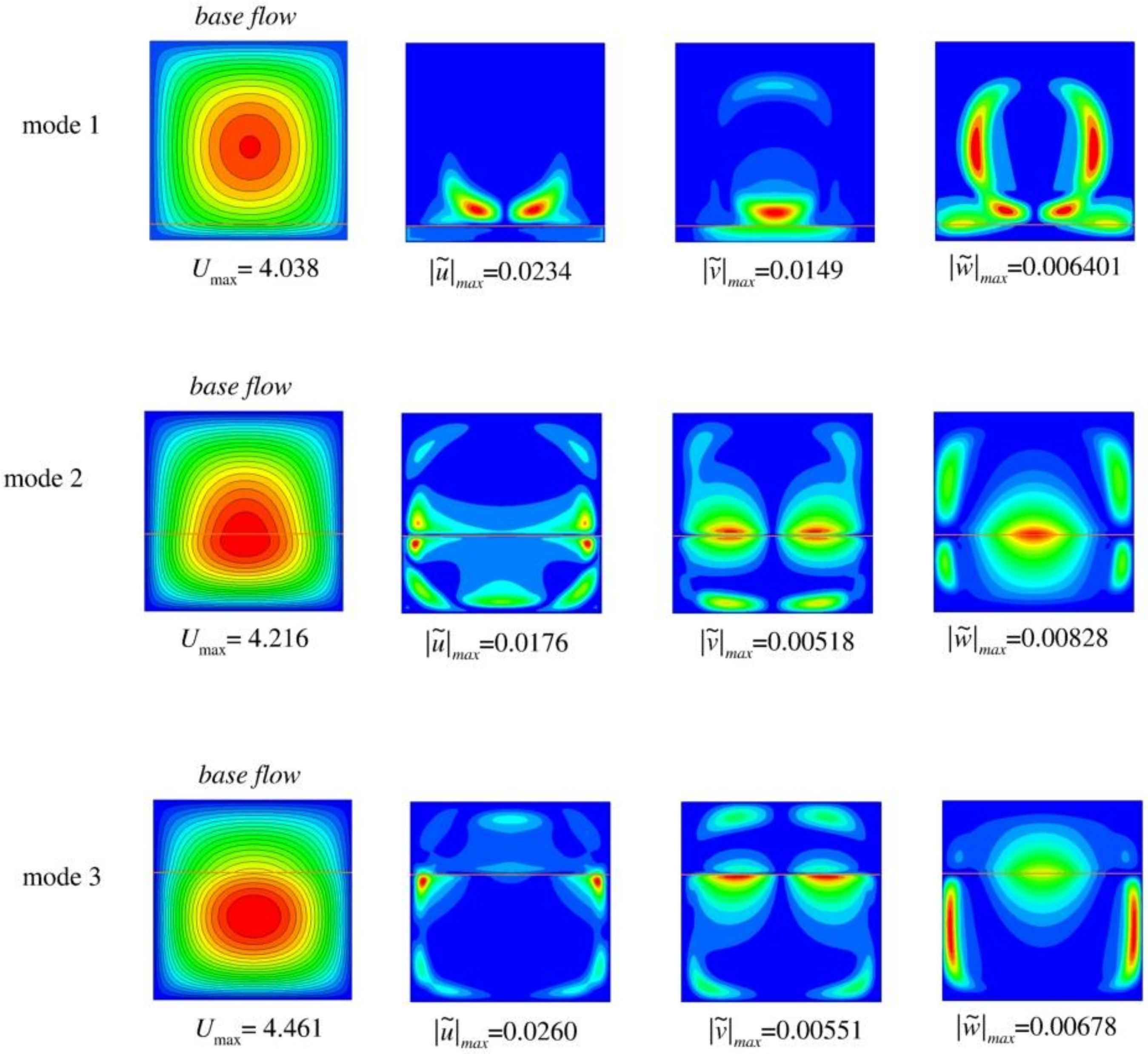

Fig. 7. The base flow (left frames) and amplitudes of the velocity perturbations typical for the three instability modes depicted in Fig. 5.2.1. Mode 1: $U_{1s} = 0.005\, m/sec\,, U_{2s,cr} = 0.159\, m/sec\,, h_1 = 0.079$; Mode 2: $U_{1s} = 0.05\, m/sec\,, U_{2s,cr} = 0.0726\, m/sec\,, h_1 = 0.389$; Mode 3: $U_{1s} = 0.13\, m/sec\,, U_{2s,cr} = 0.0409\, m/sec\,, h_1 = 0.640$.

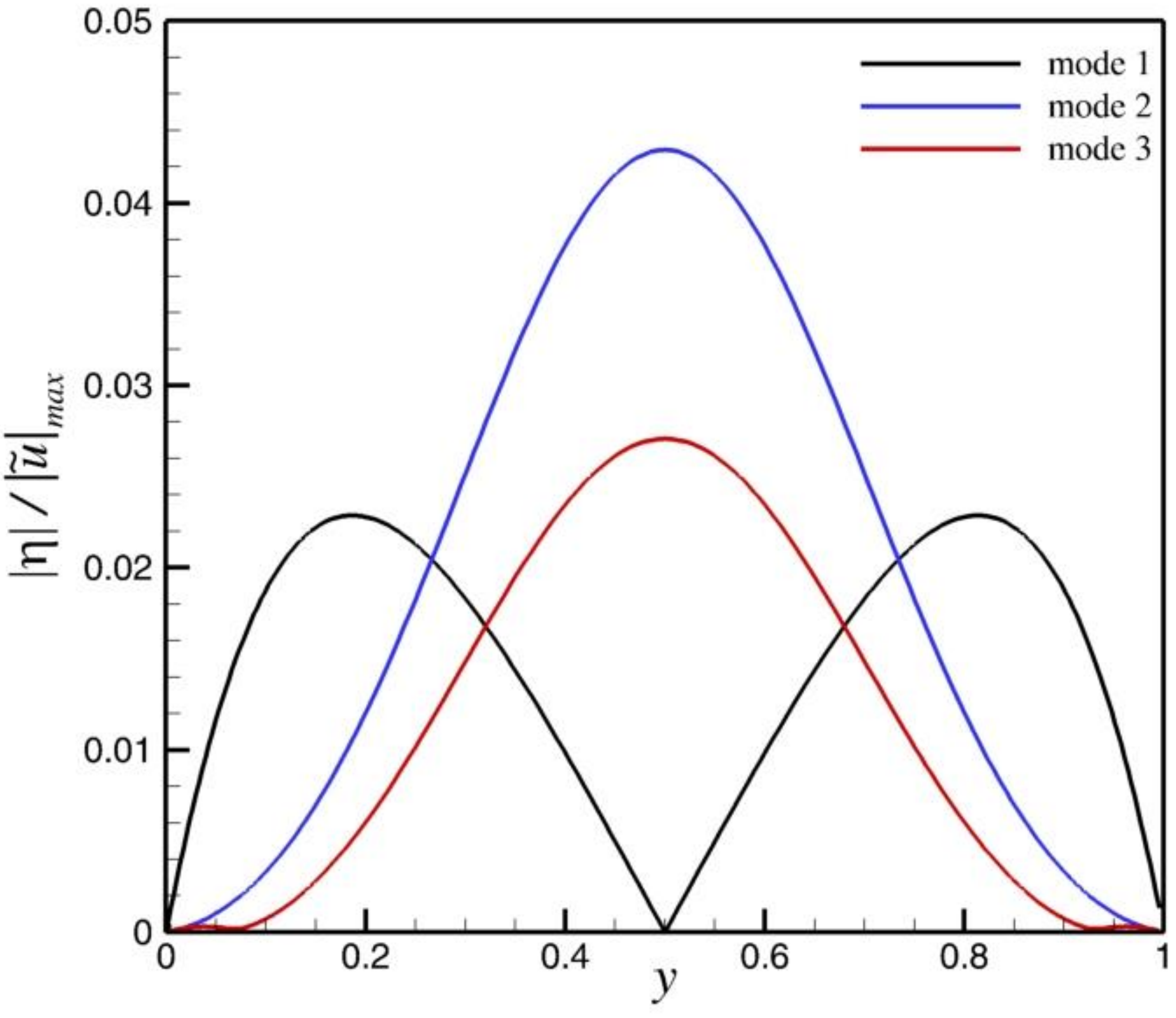

Fig. 8. Span-wise variation of the interface deformation amplitude for the three most unstable modes depicted in Fig. 5.2.2.

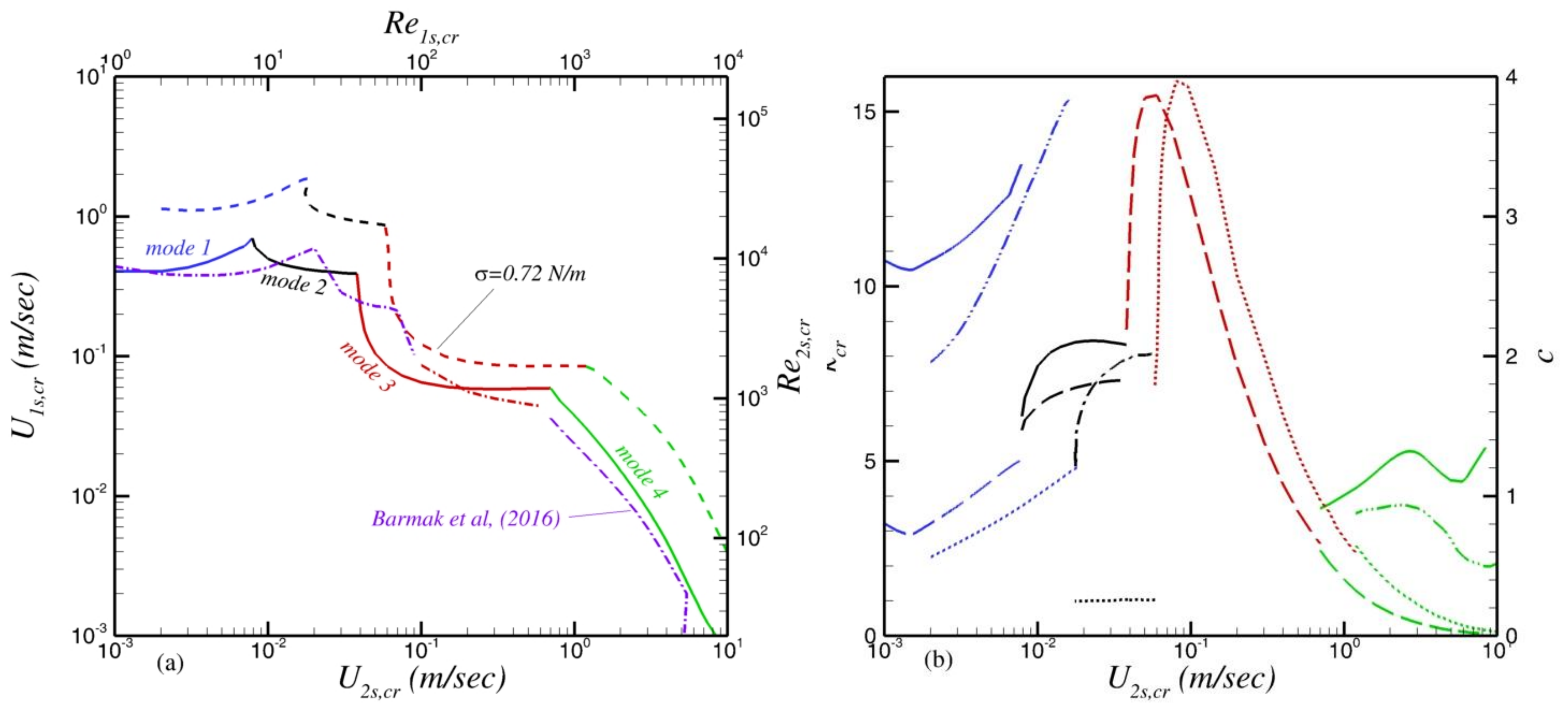

Fig. 9. (a) Stability map and (b) variation of the dimensionless critical wavenumber (solid lines) and the wave speed (dash lines) for an air – water flow in a square channel ($H$=2cm , $A$=1) . $\rho_1 = 1000\,kg/m^3$, $\mu_1 = 10^{-3}\,kg/m\cdot sec$ , $\rho_2 = 1\,kg/m^3$ , $\mu_2 = 1.8\cdot 10^{-5}\,kg/m\cdot sec, \sigma = 0.072\,N/m$ and $0.72\,N/m$ , $H = 0.02m$. Solid lines – critical superficial velocities (a) and critical wavenumbers (b) for $\sigma = 0.072\,N/m$; lines with long dashes in frame (b) – critical wave speed for $\sigma = 0.072\,N/m$; lines with short dashes in frame (a) – critical superficial velocities for $\sigma = 0.72\,N/m$; dash-dot-dot lines in frame (b) – critical wavenumbers for $\sigma = 0.72\,N/m$; dot lines in frame (b) – critical wave speed for for $\sigma = 0.72\,N/m$. Dash-and-dot violet line in frame (a) – results of Barmak et al (2016) for the flow between infinite plates.

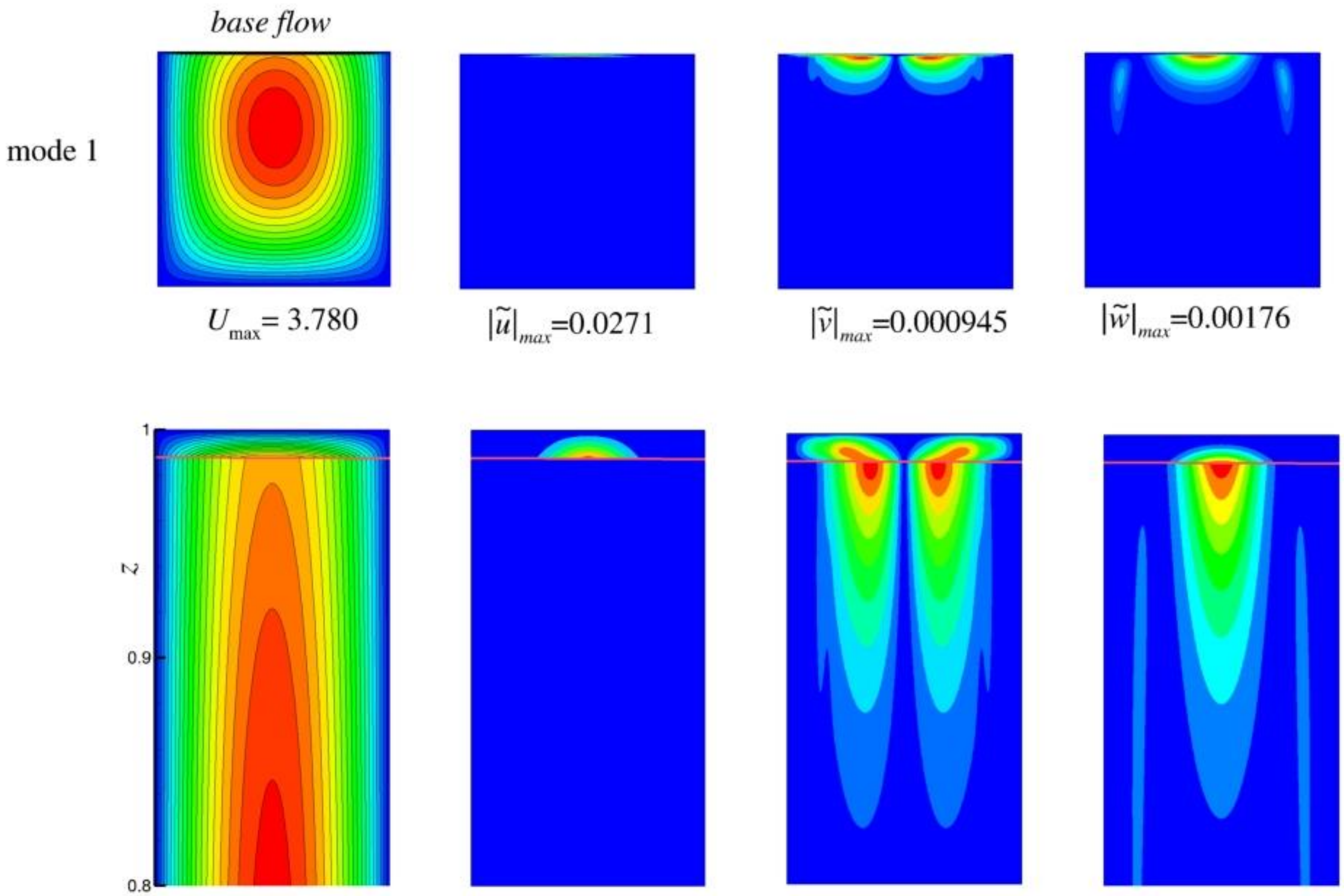

Fig. 10. The base flow (left frames) and amplitudes of the velocity perturbations typical for the instability mode 1 depicted in Fig. 5.3.1. $U_{1s} = 0.431\, m/sec$ , $U_{2s,cr} = 0.003\, m/sec$ , $h_1 = 0.988$. The base flow and the perturbations in the upper layer are zoomed in in the lower frames.

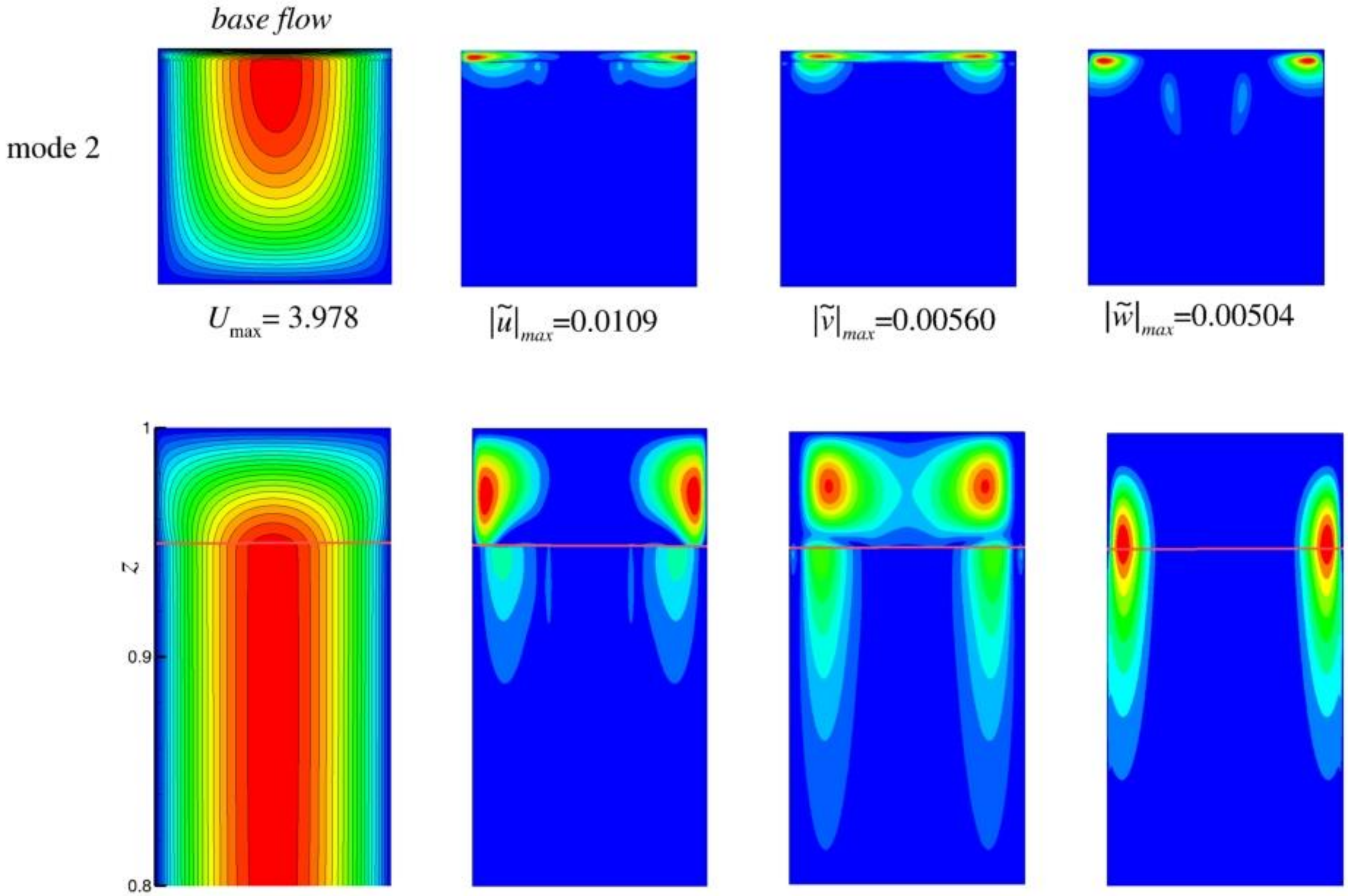

Fig. 11. The base flow (left frames) and amplitudes of the velocity perturbations typical for the instability mode 2 depicted in Fig. 5.3.1. $U_{1s} = 0.413\, m/sec$ , $U_{2s,cr} = 0.02\, m/sec$ , $h_1 = 0.948$. The base flow and the perturbations in the upper layer are zoomed in in the lower frames.

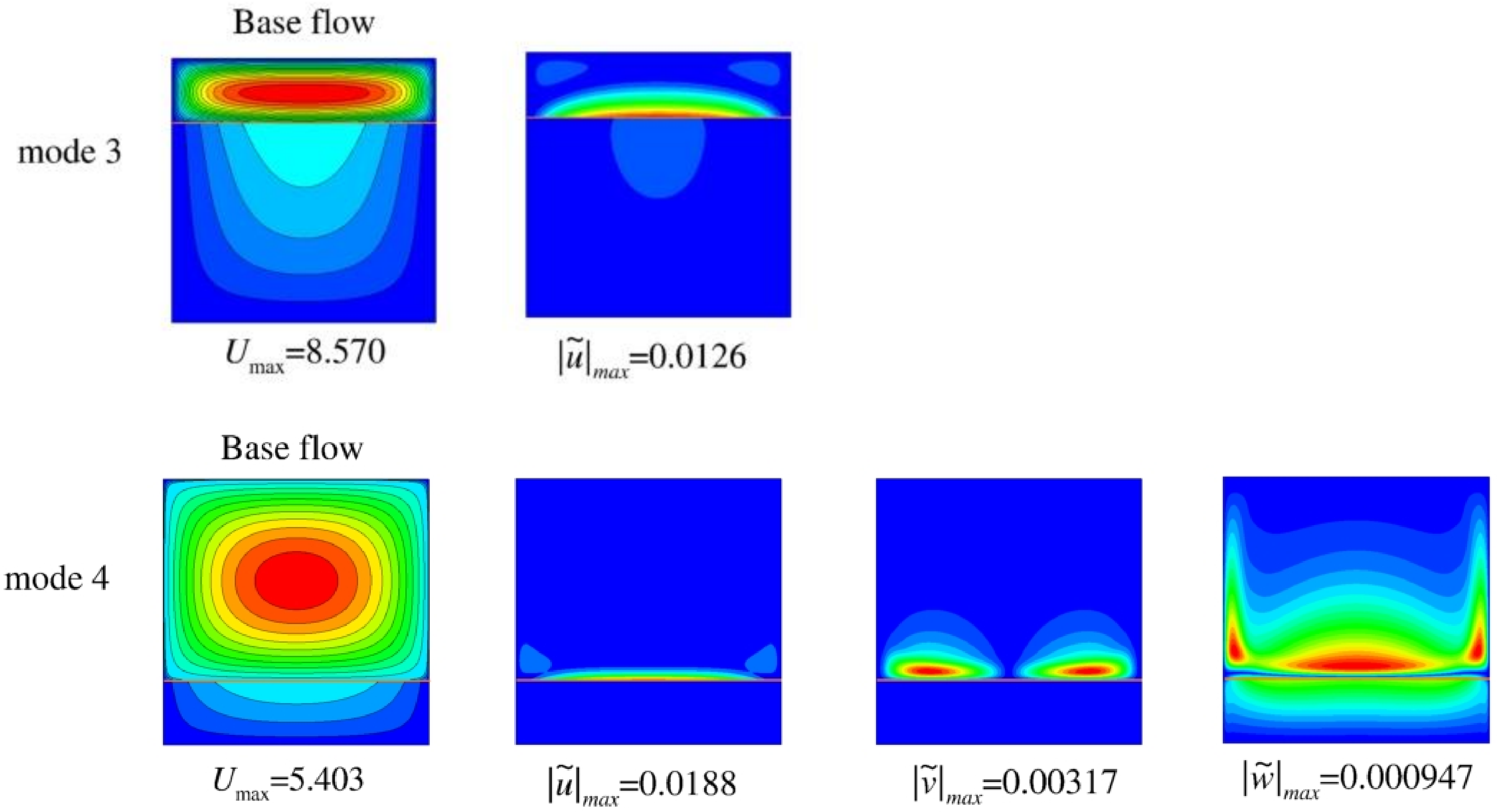

Fig. 12. The base flow (left frames) and amplitudes of the velocity perturbations typical for the instability mode 3 and 4 depicted in Fig. 5.3.1. Mode 3 is the long-wave mode: $U_{1s} = 0.651\, m/sec$, $U_{2s,cr} = 0.1\, m/sec$, $h_1 = 0.756$, perturbations $\tilde{u}$ and $\tilde{v}$ vanish; Mode 4: $U_{1s} = 0.0152\, m/sec$, $U_{2s,cr} = 2\, m/sec$, $h_1 = 0.242$

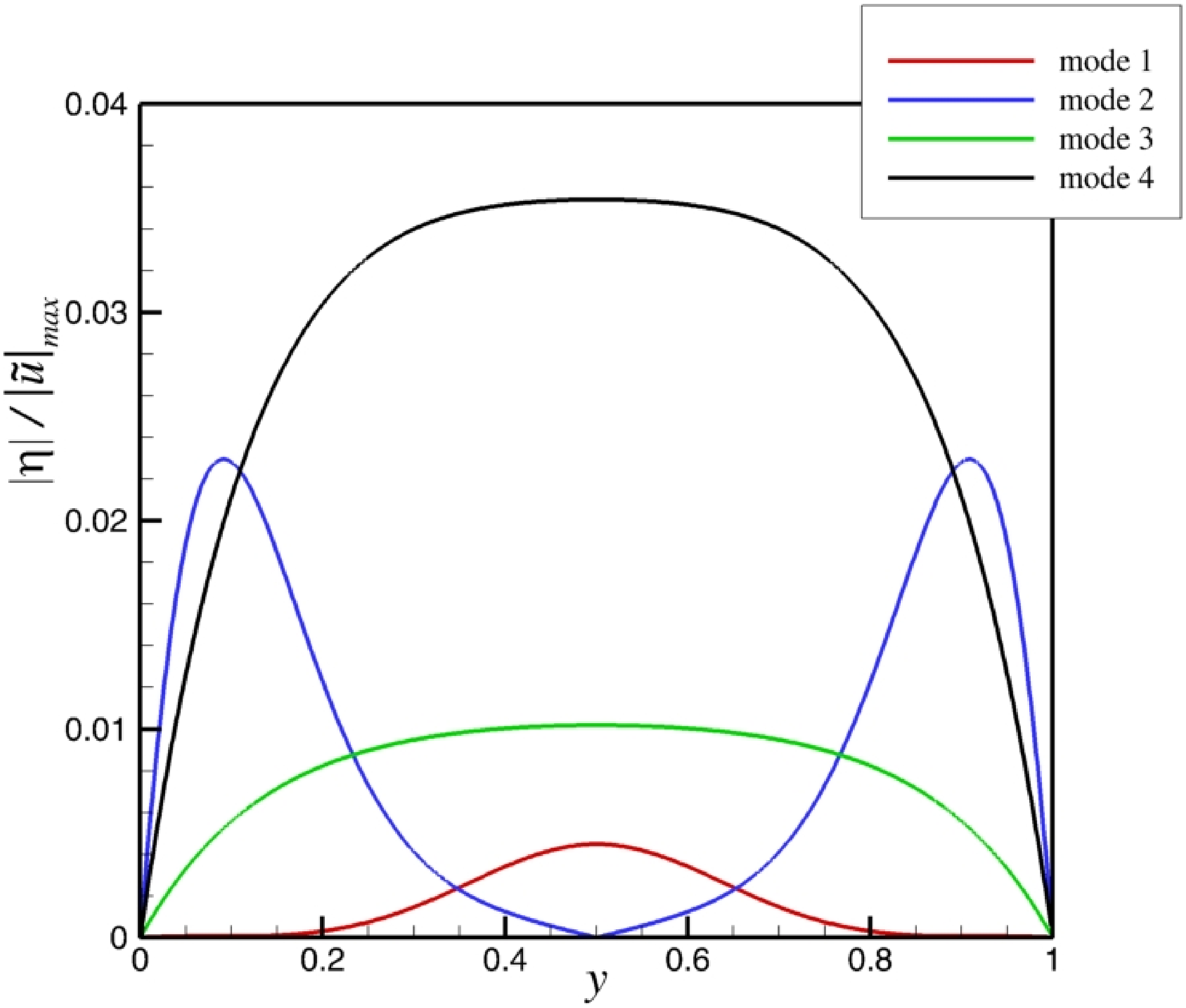

Fig. 13. Span-wise variation of the interface deformation amplitude for the three most unstable modes depicted in Fig. 5.3.1.